\documentclass[english,floatfix,twocolumn,showpacs,superscriptaddress,nofootinbib,aps,prd]{revtex4}
\usepackage[T1]{fontenc}
\usepackage{lmodern}
\usepackage{amsmath}
\usepackage{amssymb}
\usepackage{graphicx}
\usepackage{esint}
\usepackage{longtable}
\usepackage{dcolumn}
\usepackage{babel} 
\usepackage{csquotes}

\begin{document}

\title{Upper bound on the GUP parameter using the black hole shadow}

\author{Juliano C. S. Neves}
\email{nevesjcs@if.usp.br}
\affiliation{Centro de Ciências Naturais e Humanas, Universidade Federal do ABC,\\ Avenida dos Estados 5001, Santo André, 09210-580 São Paulo, Brazil}

%%%%%%%%%%%%%%%%%%%%%%%%%%%%%%%%%%%%%%%%%%%%%%%%%%%%%%%%

\begin{abstract}
An upper bound on the parameter that provides a generalized uncertainty principle  (GUP) is obtained from the black hole shadow. With the aid of a recent constraint between regular black holes and the GUP parameter, it is indicated a relation between this parameter and the deviation from circularity of the black hole shadow. In the case of the recent announcement of the M87* results from the Event Horizon Telescope collaboration, a deviation from circularity $\lesssim 10\%$ imposes a GUP parameter $\beta_0 <10^{90}$.     
\end{abstract}


\maketitle

\section{Introduction}

The Event Horizon Telescope collaboration announced the first black hole image ever captured \cite{EHT,EHT2}. By
 using interferometry, the collaboration built the shadow of the supermassive black hole (M87*) at the center of the Messier 87 galaxy. As pointed out by the collaboration, the black hole shadow is well-described by general relativity by  adopting the Kerr metric in order to interpret the phenomenon. Moreover, the supermassive M87*---whose mass is  $M=(6.5 \pm 0.7 )\times 10^{9}M_{\odot}$ and is $r_{obs}= (16.8 \pm 0.8)$Mpc distant from us---presents an
     almost circular shadow. Accordingly, the shadow of M87* indicates a deviation $\lesssim 10\%$ from circularity. Even
      with the good agreement with the Kerr metric, options are not totally ruled out. Due to the uncertainty on both the
       rotation  parameter and observation angle, options to Kerr metric are still candidates for M87* shadow \cite{Bambi}. This article explores that avenue and applies the M87* parameters to Kerr-like objects, like rotating regular black holes.   

Shadows of black holes have been studied since the pioneer work of Synge \cite{Synge}, in which the shadow of the
 Schwarz-schild  black hole was obtained. Bardeen \cite{Bardeen2} built the first shadow for a rotating black hole,
  namely,  for the Kerr metric. Since then, whether in the general relativity context or beyond, shadows of different 
  black holes have been published, like shadows for the Reissner-Nordström
    black hole \cite{Zakharov}, the Kerr-Newman black hole \cite{Vries}, black holes with a cosmological constant
     \cite{Perlick}, Kerr-Newman-NUT black hole \cite{Grenzebach}, Kerr-Newman-Kasuya black hole \cite{Ovgun}, braneworld black holes \cite{Eiroa1,Eiroa2},
      rotating wormholes \cite{Shaikh} and for regular black holes \cite{Zi_Bambi,Abdujabbarov,Amir}, which are focused
       on this article. It is worth emphasizing that the results exposed here do not consider an accretion disk around the black
        hole. Accretion disks modify the shadow's silhouette, and models that consider such influences are being developed
         \cite{Perlick2,Cunha,Roman}. The results presented in this article can be used in future researches that include accretion disks, whether made of plasma or dark matter.    
 
As an alternative to singular black holes, regular black holes are solutions of the gravitational field equations without a
 singularity inside the event horizon. Once again, Bardeen was pionner when constructed the very first regular metric that describes a regular black hole \cite{Bardeen}. The Bardeen black hole is spherical, i.e., a nonrotating black hole, and is described by a mass function that depends on the radial coordinate. Inside the event horizon, the Bardeen metric hides a de
 Sitter core instead of a Schwarzschild singularity. A de Sitter core inside the event horizon avoids either the point singularity of spherical black holes \cite{Ansoldi,Lemos_Zanchin,Dymnikova,Dymnikova2,Dymnikova3,Bronnikov,Hayward,Neves,Neves2} or the ring singularity in geometries with axial symmetry \cite{Various_axial,Various_axial2,Various_axial3,Various_axial4,Various_axial5,Neves_Saa}. As is well-known, the de
 Sitter core provides energy conditions violations in order to avoid consequences of the singularities theorems.\footnote{See Wald's book \cite{Wald} for detailed studies on the theorems.} Then, the Bardeen regular black hole violates the strong energy condition, and, according to our work \cite{Neves_Saa}, those with rotation ignore the weak energy condition.   
 
Besides a mass function that depends on the radial coordinate, regular black holes possess a mass function that depends on the parameter related directly to the regular geometry or the absence of a singularity. In Ayón-Beato and Garcia's work
 \cite{Beato}, such a parameter is a charge, and the Bardeen regular black hole is conceived of as a solution of general
  relativity coupled to a nonlinear electrodynamics for those authors. However, recently, our work \cite{Maluf_Neves2}
   suggested another interpretation to the Bardeen metric. By using a generalized uncertainty principle (GUP), we
    computed quantum corrections to the Hawking temperature for the Schwarzschild black hole. Thus, we pointed out
     that, at
     second order, Bardeen's regular black hole may be view as a quantum-corrected  Sch-warzschild black hole. This new
      interpretation led to a constraint between a metric parameter, namely, the parameter in the mass function that produces
       regularity, and the GUP parameter. With the aid of such a relation, the GUP parameter will be estimated using the
        black hole shadow.  The regular black holes used in this shadow calculation come from a class of rotating regular
         black holes with a cosmological constant published in our article \cite{Neves_Saa}.
  
GUPs appear, for example, in candidates for quantum gravity theories and in the deformed quantum mechanics,
 where the canonical commutation relations are modified in order to provide a minimal length scale and,
 consequently, its contribution to empirical results, like in the modified hy-drogen-atom spectrum
  \cite{Benczik}, Lamb shift, Landau Levels and in the
       scanning tunneling microscope  \cite{Das_Vagenas,Das_Vagenas2}.\footnote{See Ref. \cite{Tawfik}
for a review on GUPs.} The dimensionless  GUP  parameter (also called quantum gravity parameter)  deforms
 the Heisenberg relation, and with  $\beta_0=0$, where  $\beta_0$ is the mentioned parameter, Heisenberg's 
 uncertainty relation is recovered. There is a  debate on the value of  the GUP parameter \cite{Das_Vagenas,Scardigli_Casadio,Feng}. Assuming that the GUP
   parameter is   $\beta_0 \sim 1$ implies that effects of $\beta_0$ are hard or too small to be detected. But without such
    an \textit{a priori} assumption, it is possible  to obtain upper bounds on the GUP parameter by using recent
     experiments. Like Ref. \cite{Scardigli_Casadio}, where the  upper bounds were built with the aid of the light deflection
      and perihelion precession,  and Ref. \cite{Feng}, where gravitational waves  were adopted, in this work an option in
       the strong gravitational field regime (the shadow of M87*) is used in order to obtain an upper bound on $\beta_0$.
        In general, as we will see,
      gravitational systems offer worse upper bounds than quantum options, like the Lamb shift, Landau Levels or the
       scanning tunneling microscope used in Ref. \cite{Das_Vagenas}.  
  
 As I said, the deviation from circularity, reported by the collaboration, will be used in order to estimate the  GUP
  parameter. It is worth emphasizing that the class of rotating regular black holes used here and presented in Ref.
   \cite{Neves_Saa} generalizes earlier Kerr-like regular solutions because it possesses a general mass function $m(r)$ and
    a cosmological constant. The shadow of that class has its silhouette  presented here for the first time in the literature.
     With the M87* parameters, by assuming that the angle between the black hole rotation axis and the observer is 
     $\theta_{obs}=17^{\circ}$ (in agreement with observed jets supposedly aligned with the rotation axis \cite{EHT,Walker}), it is indicated the $\beta_0<10^{90}$  as upper bound on the GUP parameter. 

The structure of this paper is as follows: in Section 2, the geodesic equations for the class of  rotating regular black holes   and equations that provide the shadow's silhouette were derived. In Section 3, two shadow's observables are indicated, oblateness and root-mean-square distance from the average radius of the shadow, which gives the deviation from circularity, such that the latter was computed in order to provide an upper bound on the GUP parameter in Section 4. In Section 5, the final remarks are made.\footnote{In this work, I adopted geometric units such that $G=c=1$, where $G$ is the gravitational constant, and $c$ is the speed of light in vacuum. For the evaluation of the GUP parameter, the M87* data were used, and, consequently, $G$ and $c$ were restored.}

\section{The shadow of rotating regular black holes}
Let us obtain the shadow of rotating regular black holes with a cosmological constant in this section. Firstly, the spacetime metric and its geodesic equations are shown, then the silhouette equations are computed.  

\subsection{Spacetime metric and geodesic equations}
In  this article, shadows are obtained from a class of rotating regular black holes developed
 in Ref. \cite{Neves_Saa}. The spacetime metric of that class---using the  Boyer-Lindquist coordinates---is given by
\begin{eqnarray}
ds^{2} & = & -\frac{1}{\Sigma}\left(\Delta_{r} - \Delta_{\theta} a^{2} \sin^{2} \theta\right) dt^{2} \nonumber \\
&& - \frac{2a}{\Xi\Sigma} \left[(r^{2} + a^{2})\Delta_{\theta} - \Delta_{r}\right] \sin^{2} \theta dt d\phi \nonumber \\
&& + \frac{\Sigma}{\Delta_{r}}dr^{2} + \frac{\Sigma}{\Delta_{\theta}}d\theta^{2}\nonumber \\
& &  + \frac{\sin^{2}\theta}{\Xi^{2}\Sigma}\left[(r^{2}+a^{2})^{2}\Delta_{\theta}-\Delta_{r}a^{2}\sin^{2}\theta\right] d\phi^{2},
\label{Boyer-Lindquist}
\end{eqnarray}
where
\begin{equation}
\Delta_{\theta} = 1 + \frac{\Lambda}{3} a^{2} \cos^{2} \theta, \hspace{0.25cm} \Sigma = r^{2} + a^{2} \cos^{2} \theta, \hspace{0.25cm}\Xi = 1 + \frac{\Lambda}{3} a^{2},
\label{definitions}
\end{equation}
and
\begin{equation}
\Delta_{r} = (r^{2} + a^{2})\left( 1 - \frac{\Lambda}{3}r^{2}\right) - 2m(r)r.
\label{Delta_r}
\end{equation}
The constant $\Lambda$ is the cosmological constant, $a$ is the rotation parameter and, for the class of regular black holes studied here, the black hole mass depends on the radial coordinate $r$, namely,
\begin{equation}
m(r)=M\left(1+\left(\frac{r_{0}}{r}\right)^{q}\right)^{-\frac{3}{q}}.
\label{Mass_term}
\end{equation}
The mass function (\ref{Mass_term}) provides black holes without a singularity.\footnote{Aspects of regular black holes
 with the mass function (\ref{Mass_term}) were studied: thermodynamics in Ref \cite{Maluf_Neves1}, accretion of
  perfect fluids in Ref. \cite{Neves_Saa2}, and cosmic strings in Ref. \cite{Ceren}.} For different values of the integer $q$, we have well-known regular black holes. For example, for the spherical case, $q=2$ provides the Bardeen black hole
   \cite{Bardeen}, and $q=3$ produces the Hayward regular metric \cite{Hayward}, but the mass function also gives
    axisymmetric regular black holes or Kerr-like black holes. The parameter $M$ is the mass parameter (for large values of $r$, $m(r)\sim M$), and $r_0$ is a length parameter that provides regular metrics, parameter conceived of as a microscopical constant related to both the GUP parameter and the Planck length according to our work
 Ref. \cite{Maluf_Neves2} (in Section 4 such a parameter will be briefly discussed).

Due to cosmological observations \cite{Planck}, I will focus on the positive cosmological constant in this article. In this case, regarding $r_0\ll M$, the function $\Delta_r$ provides three roots: the inner horizon $r_-$, the event horizon $r_+$, and the cosmological horizon $r_{++}$.  Then, the spacetime structure reads
\begin{equation}
r_-<r_+<r_{++}.
\label{Structure}
\end{equation} 

In order to construct the black hole shadows of the class of rotating regular black holes given by
 Eq. (\ref{Boyer-Lindquist}), the geodesic equations are needed. Geodesics for the Kerr metric were obtained by Carter
  \cite{Carter}, who showed the separability of the geodesic equations. Carter argued that a test particle in the Kerr
   spacetime  possesses four constants of motion along geodesics. Accordingly, one has the two Killing  vector fields $\xi_t$ and $\xi_\phi$ with their respective constants, the mass of the test particle and the fourth constant, which is called Carter
    constant (indicated by $K$). Both the Kerr-anti-de Sitter and Kerr-de Sitter spacetimes present those same four constants  for test particles along geodesics, whether in the general relativity realm \cite{Hackmann} or in the brane world, as we  indicated in Ref. \cite{Neves_Molina}. The class of rotating regular black holes given by (\ref{Boyer-Lindquist}) also presents these constants.     

The geodesic equations for the spacetime (\ref{Boyer-Lindquist}) are obtained from the Hamilton-Jacobi equation:
\begin{equation}
\frac{\partial S}{\partial \sigma}=\frac{1}{2}g^{\mu\nu}\frac{\partial S}{\partial x^{\mu}}\frac{\partial S}{\partial x^{\nu}},
\label{Hamilton-Jacobi}
\end{equation}
where $\sigma$ is a parameter related to the affine parameter $\tau$ (i.e., $\tau = \delta \sigma$, with $\delta$ playing the role of mass of the test particle along the geodesic). The function $S$ is the Jacobi action, which is related to the generalized momentum through 
\begin{equation}
p_\mu \equiv \frac{\partial S}{\partial x^{\mu}}.
\label{Momentum}
\end{equation}
From the two Killing vector fields $\xi_{t}$ and $\xi_{\phi}$ given by the geometry with axial symmetry (\ref{Boyer-Lindquist}), we have the particle's constants of motion $E$ and $L$, namely, energy and angular momentum, respectively,
\begin{equation}
p_{t}=-E \ \ \  \mbox{and} \ \ \  p_{\phi} = L.
\end{equation}
Following Carter, it is assumed that $S$ may be written as
\begin{equation}
S=\pm \frac{1}{2}\delta^2 \sigma-Et+L \phi+S_\theta (\theta)+S_r(r),
\label{S}
\end{equation} 
where plus and minus mean the de Sitter and anti-de Sitter cases, respectively,\footnote{From the condition $g_{\mu\nu}\dot{x}^\mu \dot{x}^\nu=\pm\delta^2$, the signal plus and minus are due to the norm of timelike vectors in de Sitter and
 anti-de Sitter spacetimes, respectively.} and expressions for $S_\theta$ and $S_r$ will be omitted here. As is known in
  general relativity, from the Lagrangian $\mathcal{L}$, the generalized momentum is defined as $p_\mu\equiv \frac{\partial \mathcal{L}}{\partial \dot{x}^\mu}= g_{\mu\nu}\dot{x}^\nu$, where dot means ordinary derivative with
   respect to the parameter $\sigma$. Therefore, from that definition and Eq.(\ref{S}), substituted into Eq.
    (\ref{Momentum}), one has the geodesic equations for the metric (\ref{Boyer-Lindquist}), which in the coordinate basis are 
\begin{eqnarray}
\Sigma\dot{t}& =& \frac{(r^{2}+a^{2})P}{\Delta_{r}}-\frac{a}{\Delta_{\theta}}\left[aE\ \sin^{2}\theta-\left(1+\frac{\Lambda}{3}a^{2}\right)L\right], \nonumber \\
\Sigma\dot{r} &=& \sqrt{\mathcal{R}}, \nonumber \\
\Sigma\dot{\theta}& =& \sqrt{\Theta},  \nonumber \\
\Sigma\dot{\phi} &= & \frac{aP}{\Delta_{r}}-\frac{1}{\Delta_{\theta}}\left[aE-\left(1+\frac{\Lambda}{3}a^{2}\right)\textrm{cosec}^{2}\theta\ L\right],
\label{Geodesics}
\end{eqnarray}
where $\Sigma$ is given by Eq. (\ref{definitions}). The functions $P$, $\mathcal{R}$ and $\Theta$ are written as
\begin{eqnarray}
P & = & \left(r^{2}+a^{2}\right)E - \left(1+\frac{\Lambda}{3}a^{2}\right)a L,
 \\
\mathcal{R} & = & P^{2}-\Delta_{r}\left(\pm\delta^{2}r^{2}+K\right) ,
 \\
\Theta & = & Q - \cos^{2}\theta \biggl[ a^{2}\left(\pm\Delta_{\theta}\delta^{2}-E^{2}\right) 
\nonumber \\
& & + \left(1+\frac{\Lambda}{3}a^{2}\right)^{2}\textrm{cosec}^{2}\theta\ L^{2} \biggl] .
\label{functions}
\end{eqnarray}
As I said, the parameter $\delta$ represents the mass of the particle along the geodesic, so $\delta=0$ in the case studied here. The constant $Q$  is related to Carter's constant $K$, that is to say,
\begin{equation}
Q = \Delta_{\theta}K-\left[\left(1+\frac{\Lambda}{3}a^{2}\right)L-aE\right]^{2},
\end{equation}
which vanishes for equatorial orbits. 

\subsection{The shadow's silhouette}
In particular, for the shadow phenomenon, only photons orbits, or null geodesics, will be adopted ($\delta=0$). And as the cosmological context is considered in this article, the cosmological constant will be assumed positive, i.e., $\Lambda >0$. Following \cite{Grenzebach}, two new parameters are defined as
\begin{equation}
\xi=\frac{L}{E} \ \ \ \ \mbox{and} \ \ \ \ \eta=\frac{K}{E^2},
\label{Constants}
\end{equation}
which are constants in the shadow's silhouette. Such a silhouette is given by unstable photon orbits with $r=r_p$ constant outside the event horizon, i.e., $r_p>r_+$. In such orbits, photons may either fall into the black hole or escape to the observer position. Thus, according to geodesic equation $\dot{r}$, we have $\mathcal{R}(r_p)=\mathcal{R}'(r_p)=0$ in order to provide the unstable orbits (the symbol $'$ means derivative with respect to $r$). Typically, for a black hole with rotation  $r_p=r_{p-}$ and $r_p=r_{p+}$ (with $r_{p-}\leq r_{p+}$), i.e., there are both a minimum and a maximum value for $r_p$, and the edge of the black hole shadow should be built for those values of $r_p$. That is, as we will see, the left and the right sides of the shadow can be different for rotating black holes due to the spacetime dragging. On the other hand, for the Schwarzschild black hole $r_{p-}=r_{p+}=3M$ (in the Schwarzschild case, $M$ is the black hole mass or the Arnowitt-Deser-Misner mass), and the shadow is perfectly symmetrical.  The equations that involve $\mathcal{R}(r_p)$ and its derivative lead to
\begin{equation}
\eta(r_p)=\frac{16r_p^2 \Delta_r(r_p)}{\Delta'_{r} (r_p)^2},
\end{equation}
and
\begin{equation}
\xi (r_p)=\frac{(r_p^2+a^2)\Delta_r'(r_p)-4r_p\Delta_r(r_p)}{\Xi a \Delta'_r(r_p)}.
\end{equation}
As we will see, these conserved quantities are part of the equations that \enquote{draw} the shadow.

\begin{figure}
\begin{centering}
\includegraphics[scale=0.55]{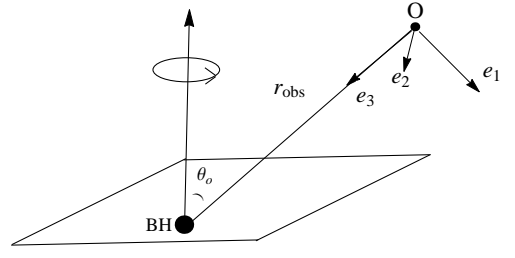}
\includegraphics[scale=0.35]{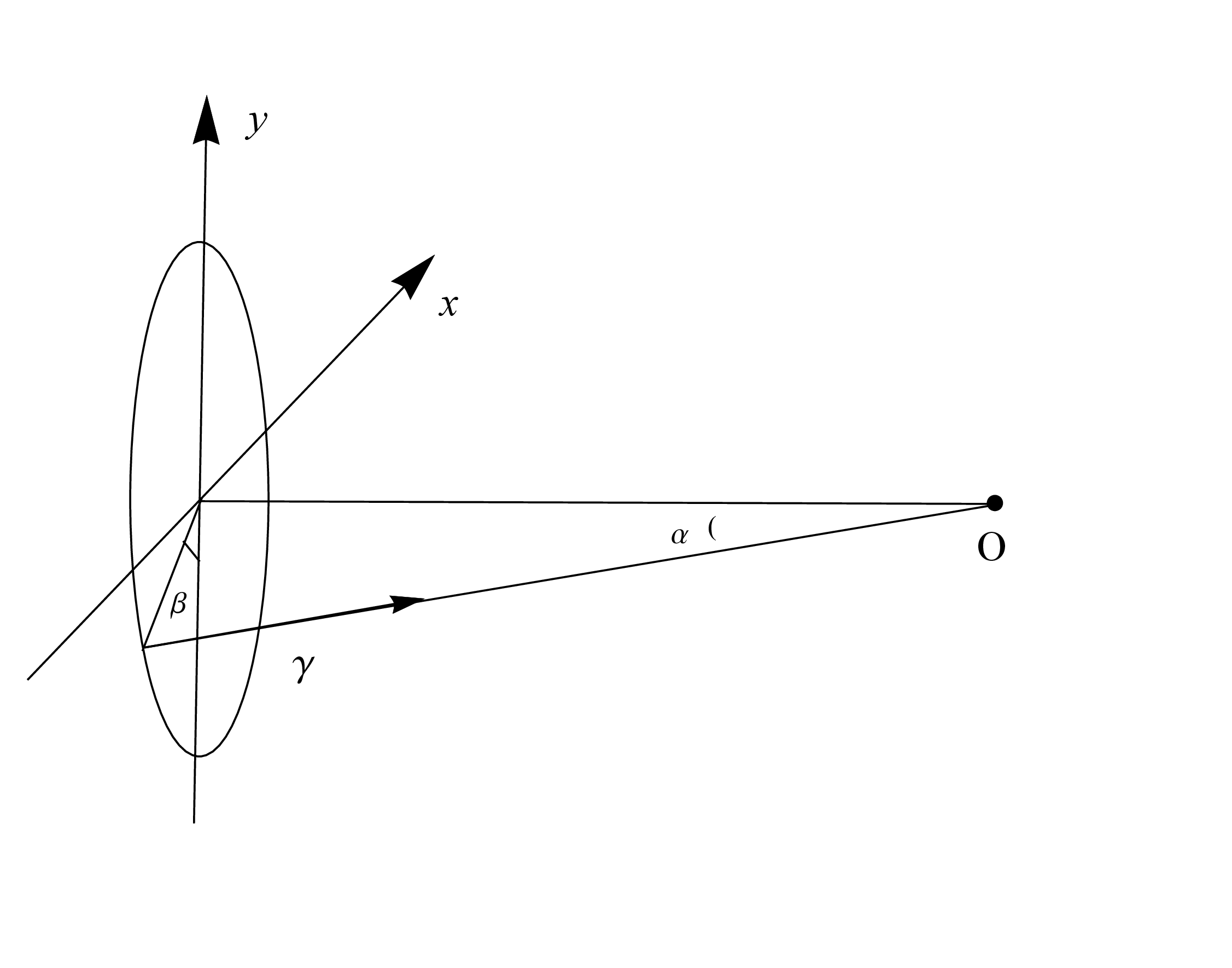}
\par\end{centering}
\caption{The shadow phenomenon observed by O at $r=r_{obs}$ using the tetrad $e_a^{\mu}$ (on the top). The celestial coordinates $\alpha$ and $\beta$ describe the null congruence $\gamma$ from the shadow's silhouette (on the bottom).}
\label{Phenomenon}
\end{figure}  

It is worth emphasizing that the metric (\ref{Boyer-Lindquist}) is not necessarily asymptotically flat, it is either
 asymptotically de Sitter or anti-de Sitter for $\Lambda \neq 0$. Thus, a given observer is not at infinity describing the
  black hole shadow, indeed he/she is in the \textit{domain of outer communication}, which is the region defined in
   between the event horizon $r_+$ and the cosmological horizon $r_{++}$ in the de Sitter case. Therefore, the observer
    will be located at the point with coordinates ($r_{obs},\theta_o$), according to Fig. \ref{Phenomenon}. In terms of the
     observer position, because of the axial symmetry, the black hole shadow depends only on the radial and
      polar coordinates. The coordinate $\theta_o$ stands for the observer angle, which is the position in which the shadow
       is observed in relation to the rotation axis, and the parameter $r_{obs}$ is the distance from the black hole. Following
       \cite{Grenzebach}, it is adopted the orthonormal tetrad $e_a^\mu=(e_0^\mu,e_1^\mu,e_2^\mu,e_3^\mu)$ in order
        to describe the shadow silhouette. That is, the null congruences coming from $r_{p-}\leq r \leq r_{p+}$ reaches the
         observer and are projected onto the tetrad, then the shadow phenomenon is described by using $e_a^\mu$. Such a tetrad is written with the aid of the coordinate basis vectors $(\partial_t,\partial_r, \partial_\theta, \partial_\phi)$, therefore we have
\begin{eqnarray}
e_0&=&\frac{(r^2+a^2)\partial_t + a\Xi \partial_\phi}{\sqrt{\Delta_r\Sigma}}\bigg\vert_{(r_{obs},\theta_o)}, \\
e_1&=&\sqrt{\frac{\Delta_\theta}{\Sigma}}\partial_\theta\bigg\vert_{(r_{obs},\theta_o)},\\
e_2&=&-\frac{\Xi\partial_\phi+a\sin^2\theta\partial_t}{\sqrt{\Delta_\theta\Sigma}\sin\theta}\bigg\vert_{(r_{obs},\theta_o)}, \\
e_3&=&-\sqrt{\frac{\Delta_r}{\Sigma}}\partial_r\bigg\vert_{(r_{obs},\theta_o)}.
\end{eqnarray}
According to Grenzebach \textit{et al.} \cite{Grenzebach}, the direction of $e_3$ points toward the black hole (see  Fig.
 \ref{Phenomenon}), and $e_0$ is the observer's four-velocity, then in this case the observer is not necessarily at rest. The
  tetrad is chosen such that $e_0\pm e_3$ is tangential to the principal null congruence direction for a metric like
   (\ref{Boyer-Lindquist}).

\begin{figure*}
\begin{centering}
\includegraphics[scale=0.45]{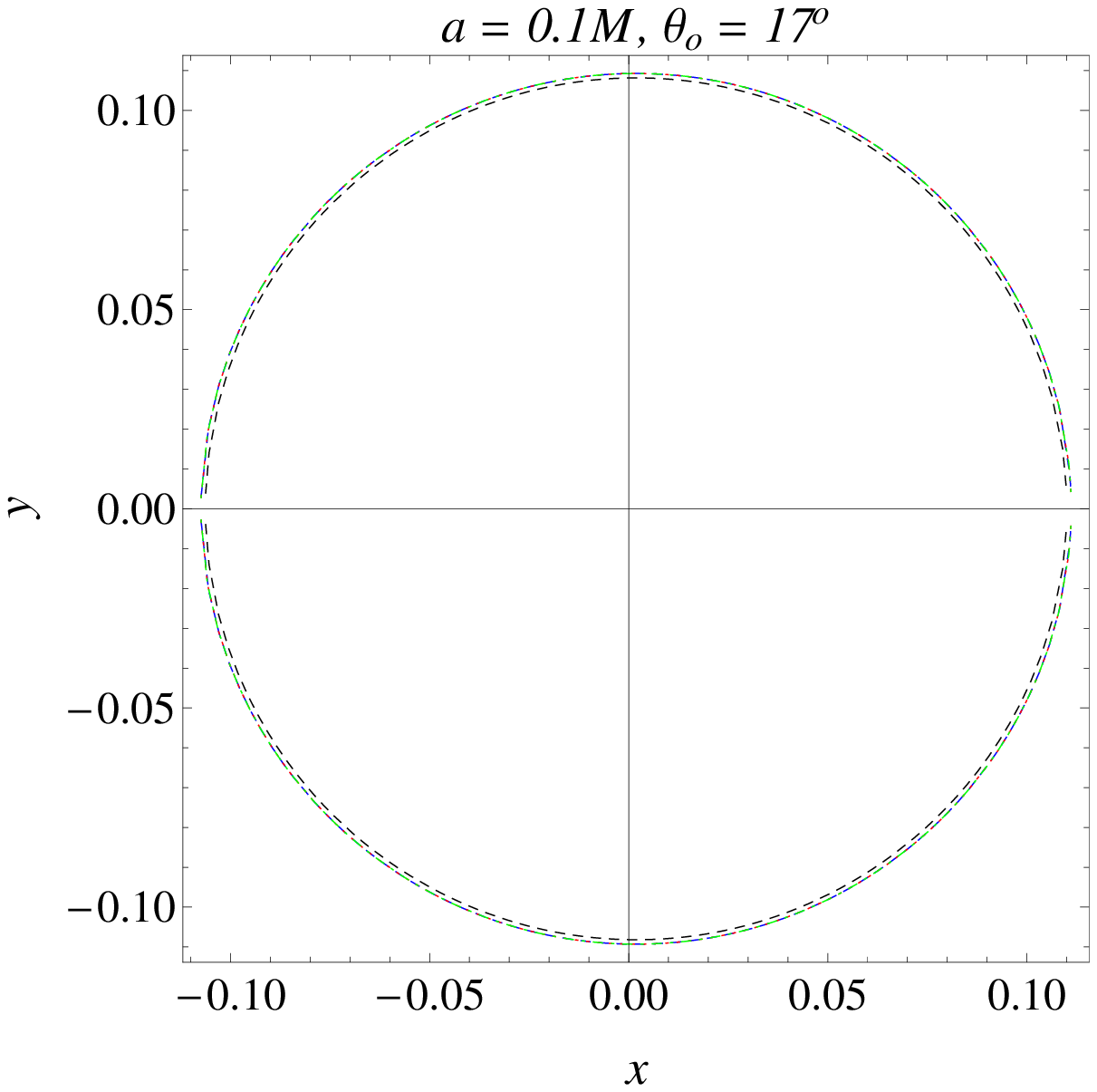}\includegraphics[scale=0.45]{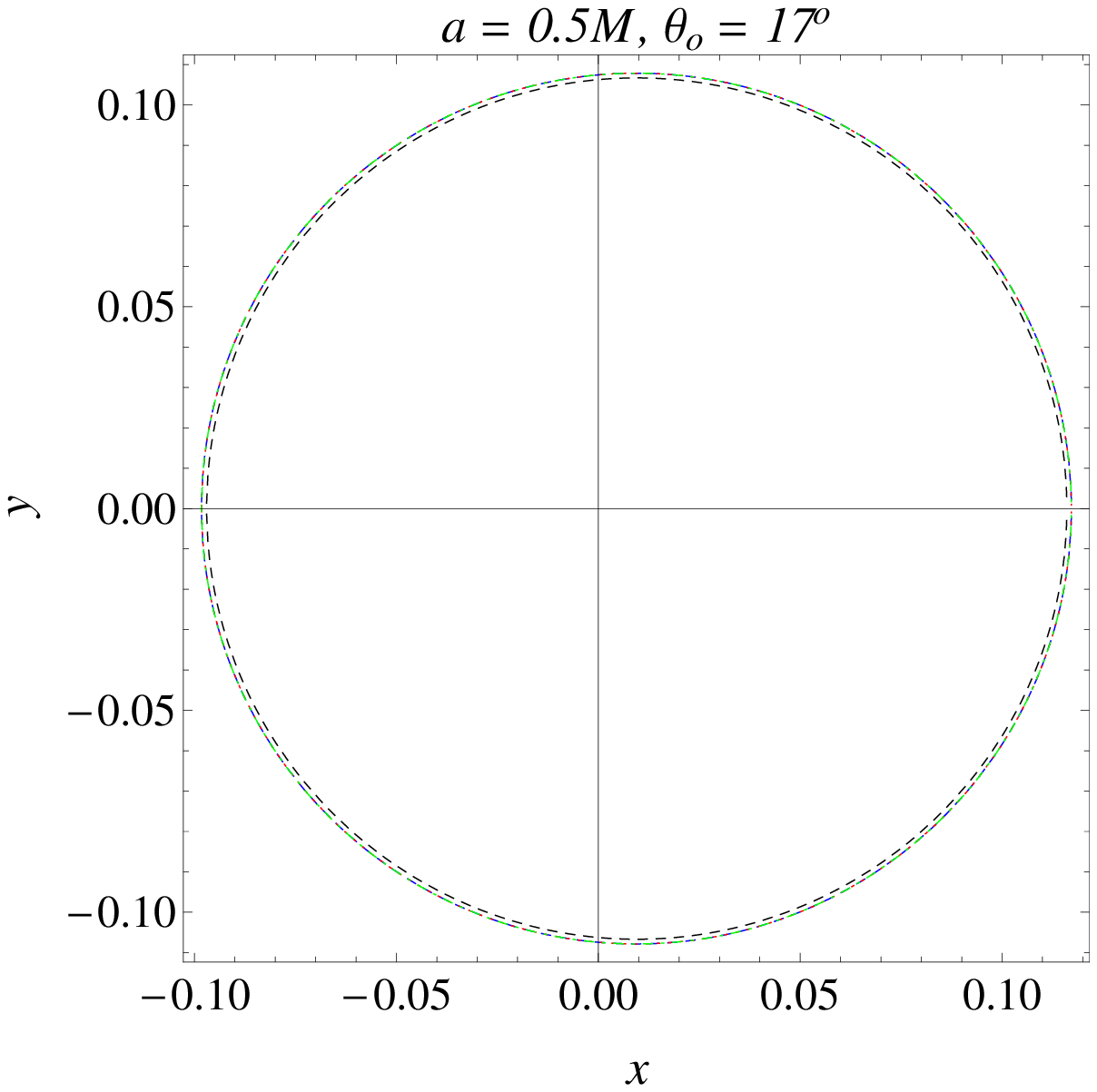}\includegraphics[scale=0.45]{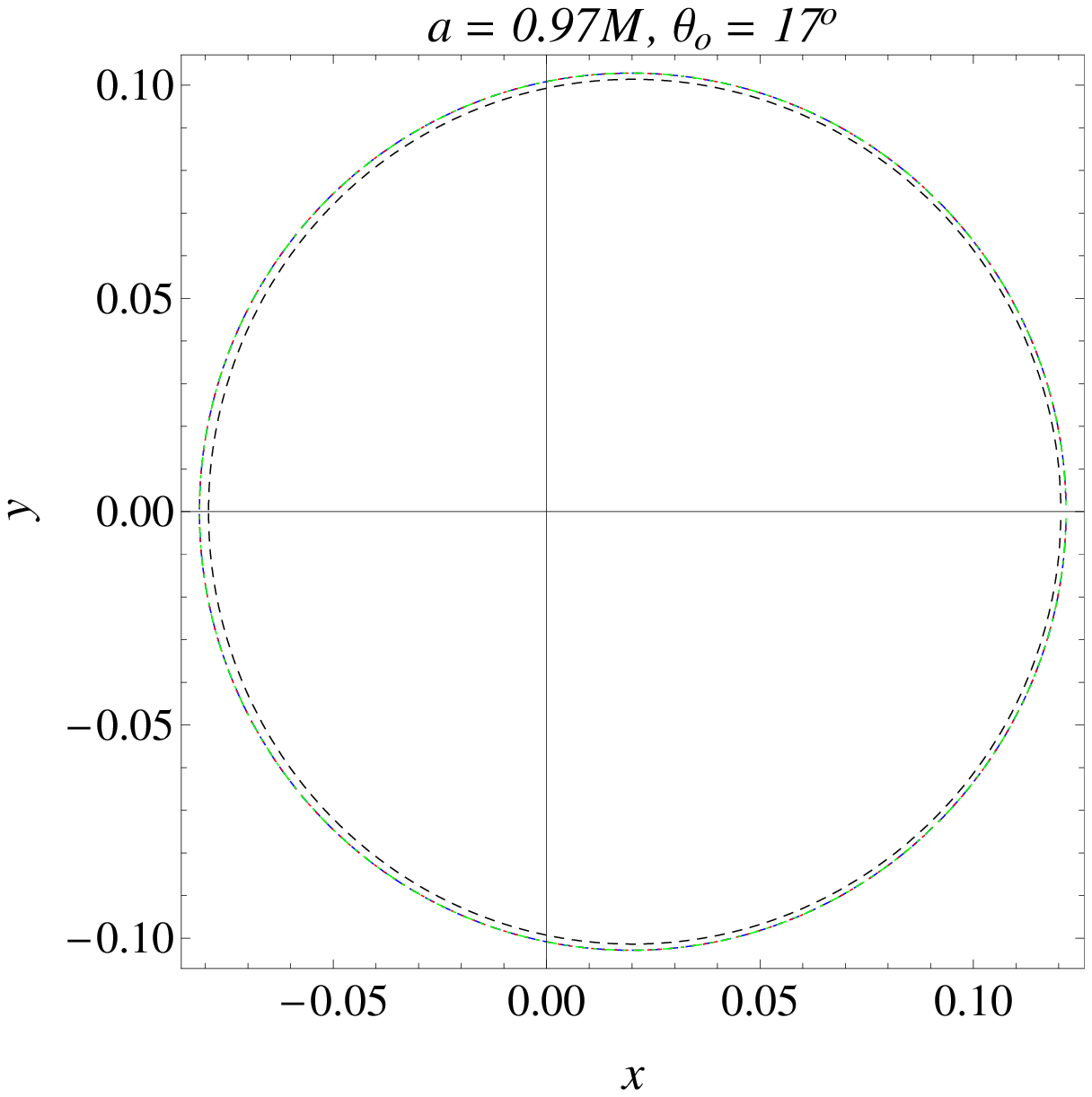}
\includegraphics[scale=0.47]{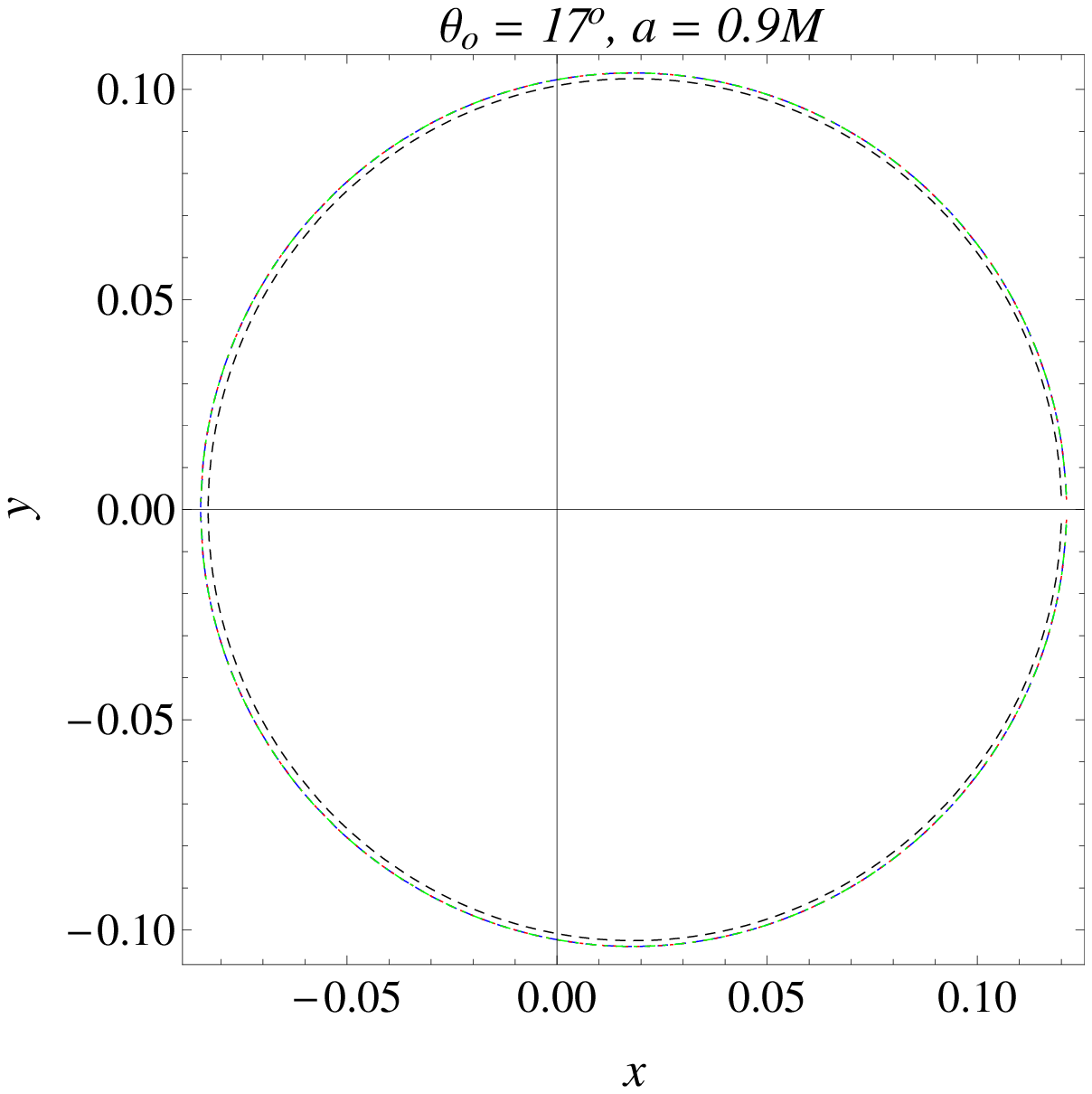}\includegraphics[scale=0.45]{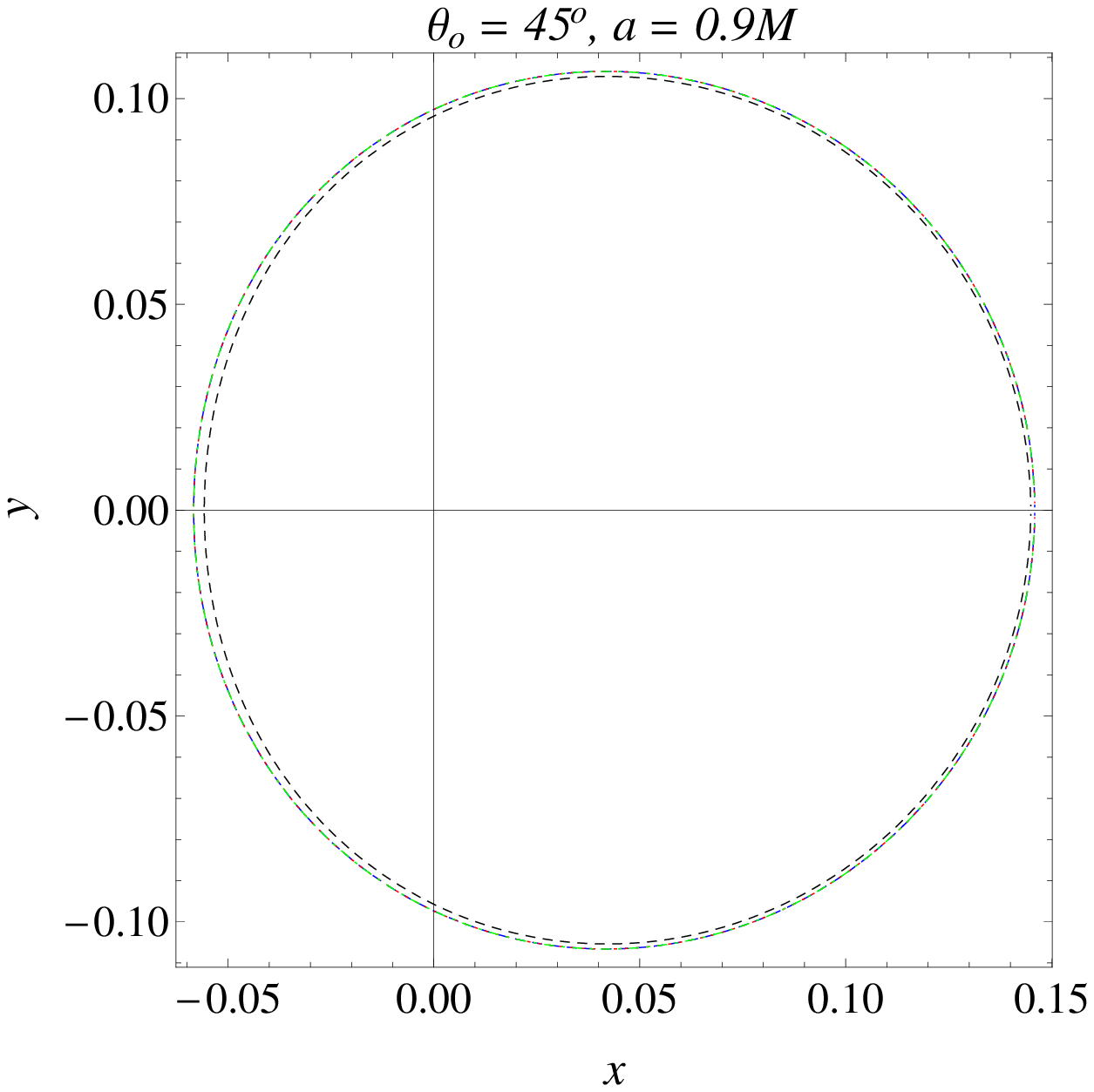}\includegraphics[scale=0.44]{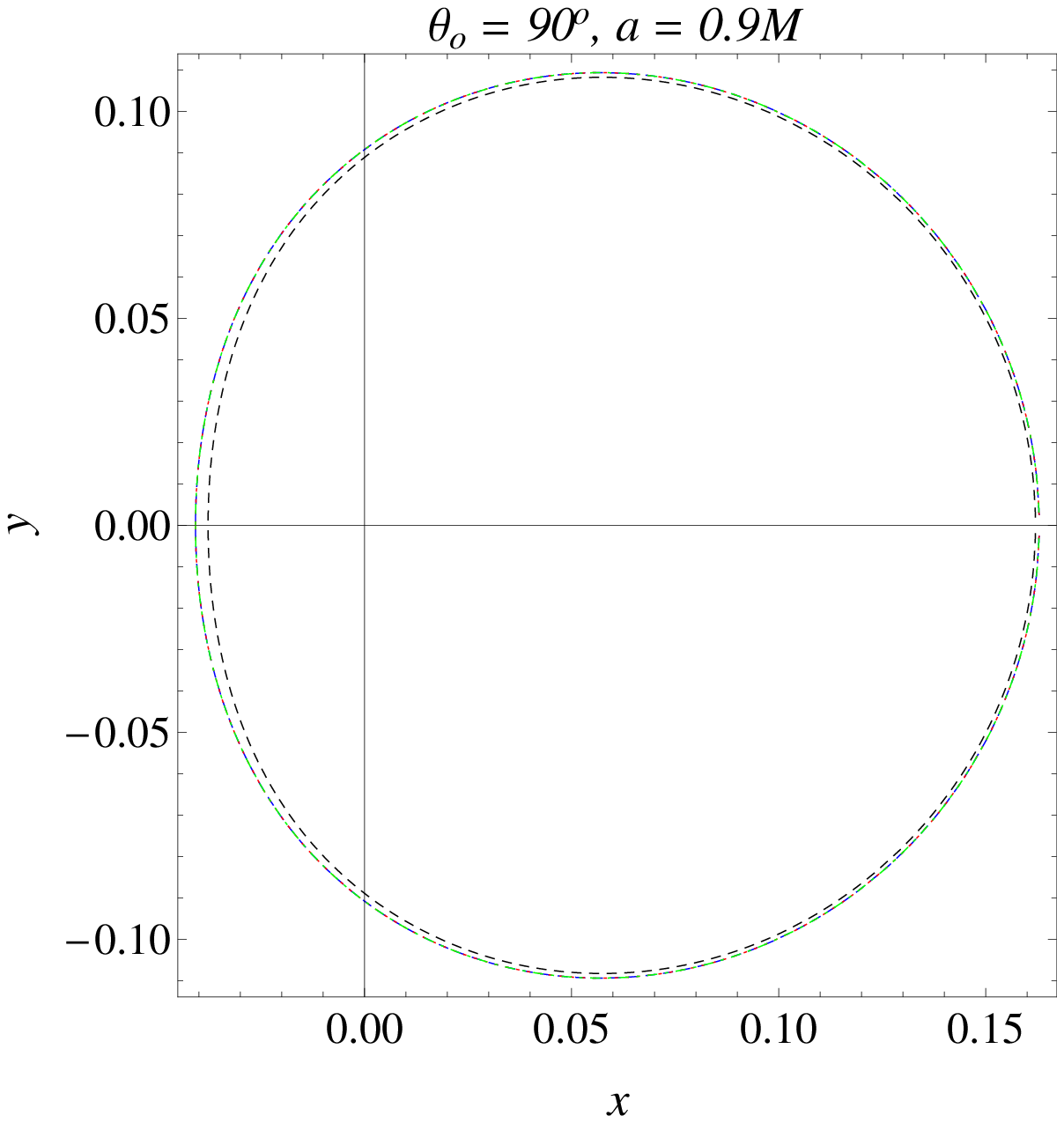}
\par\end{centering}
\caption{Shadows for the class of rotating regular black holes given by Eq. (\ref{Boyer-Lindquist}). Different values of
 $q$ in the mass function (\ref{Mass_term}) produce tiny differences in the silhouettes. For $q=1$, we have smaller
  shadows. The parameters $\Lambda=10^{-3}M$, $r_0=10^{-2}M$, $r_{obs}=35M$, and $M=1$ are adopted in these
   graphics.}
\label{Shadow1}
\end{figure*}

The null congruence, that is to say, the light rays that come from the region defined by $r_p$, are conceived of as curves $\gamma$ such that their tangent vectors are given by 
\begin{equation}
\dot{\gamma}=\dot{t}\partial_t+\dot{r}\partial_r+\dot{\theta}\partial_\theta+\dot{\phi}\partial_\phi
\label{Gamma}
\end{equation}
 in the coordinate basis. According to Ref. \cite{Grenzebach}, at the observer position, $\dot{\gamma}$ is described by the tetrad $e_a^\mu$, i.e.,
\begin{equation}
\dot{\gamma}=\zeta \left(-e_0+\sin\alpha\cos\beta e_1+\sin\alpha\sin\beta e_2+\cos \alpha e_3  \right),
\label{Gamma_dot}
\end{equation}
where the new angles $\alpha$ and $\beta$ are celestial coordinates as indicated in Fig. \ref{Phenomenon}. The description of the black hole shadow is made by using these coordinates. The factor $\zeta$ is obtained from Eqs. (\ref{Gamma})-(\ref{Gamma_dot}) and the tetrad equations, such that
\begin{equation}
\zeta=-\frac{(r^2+a^2)E-a\Xi L}{\sqrt{\Delta_r\Sigma}} \bigg\vert_{(r_{obs},\theta_o)}.
\end{equation}  
From the factor $\zeta$, by comparing terms in Eqs. (\ref{Gamma}) and (\ref{Gamma_dot}) with the aid of the equations for the tetrad, the celestial coordinates are straightforwardly written as
\begin{equation}
\cos \alpha = \frac{\Sigma\dot{r}}{(r^2+a^2)E-a\Xi L} \bigg\vert_{(r_{obs},\theta_o)},
\label{Cos}
\end{equation}
\begin{equation}
\sin \beta = \frac{\sqrt{\Delta_\theta}\sin \theta}{\sqrt{\Delta_r}\Xi \sin\alpha}\left(\frac{\Delta_r\Sigma\dot{\phi}}{(r^2+a^2)E-a\Xi L}-a\Xi \right) \bigg\vert_{(r_{obs},\theta_o)}.
\label{Sin}
\end{equation}
Then, using the geodesic equations for $\dot{r}$ and $\dot{\phi}$, substituting them into Eqs. (\ref{Cos})-(\ref{Sin}), we obtain  simple expressions for $\alpha$ and $\beta$, namely,
\begin{eqnarray}
\sin \alpha&=&\frac{\sqrt{\Delta_r \eta(r_p)}}{(r^2+a^2)-a\Xi \xi(r_p)} \bigg\vert_{(r_{obs},\theta_o)}, \\
\sin \beta &=& \frac{\left( \Xi \xi(r_p)\csc^2 \theta-a \right)\sin \theta}{\sqrt{\Delta_\theta \eta(r_p)}} \bigg\vert_{(r_{obs},\theta_o)}.
\end{eqnarray}
As we can see, the equations that describe the celestial coordinates depend on the observer position $(r_{obs},\theta_o)$ and parameters of the photon orbits, also known as photon \enquote{sphere}. 

\begin{figure*}
\begin{centering}
\includegraphics[scale=0.45]{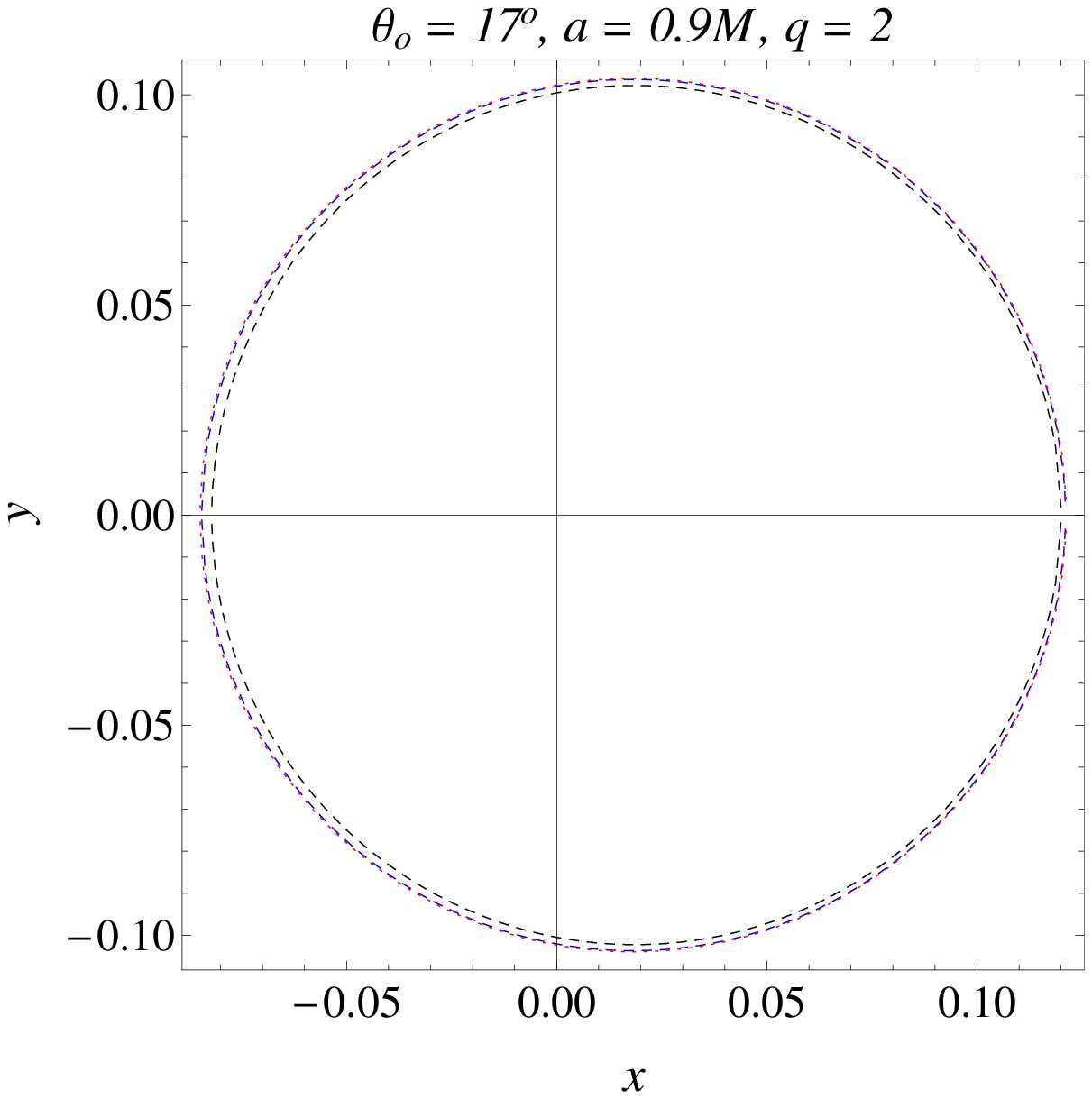}\includegraphics[scale=0.45]{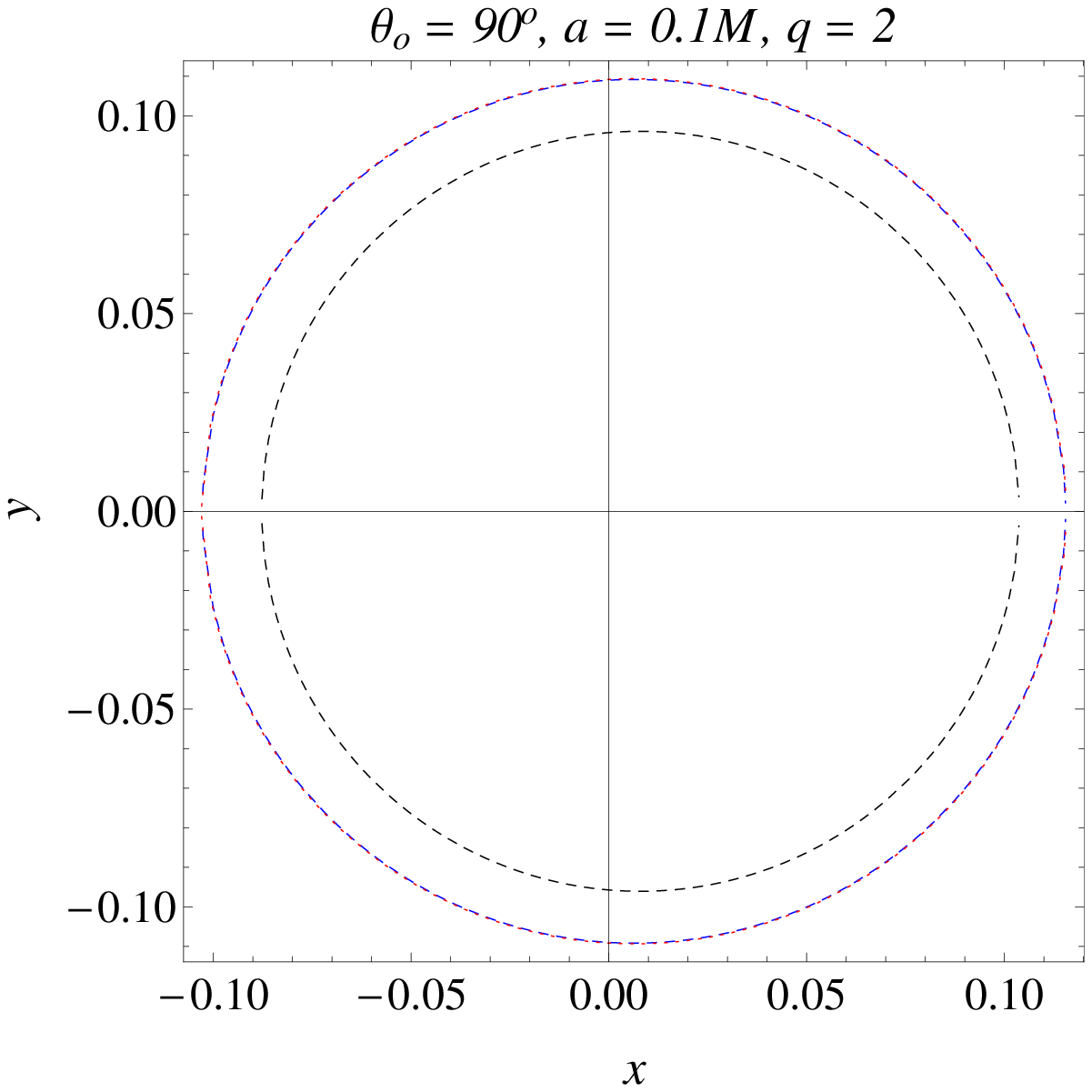}\includegraphics[scale=0.42]{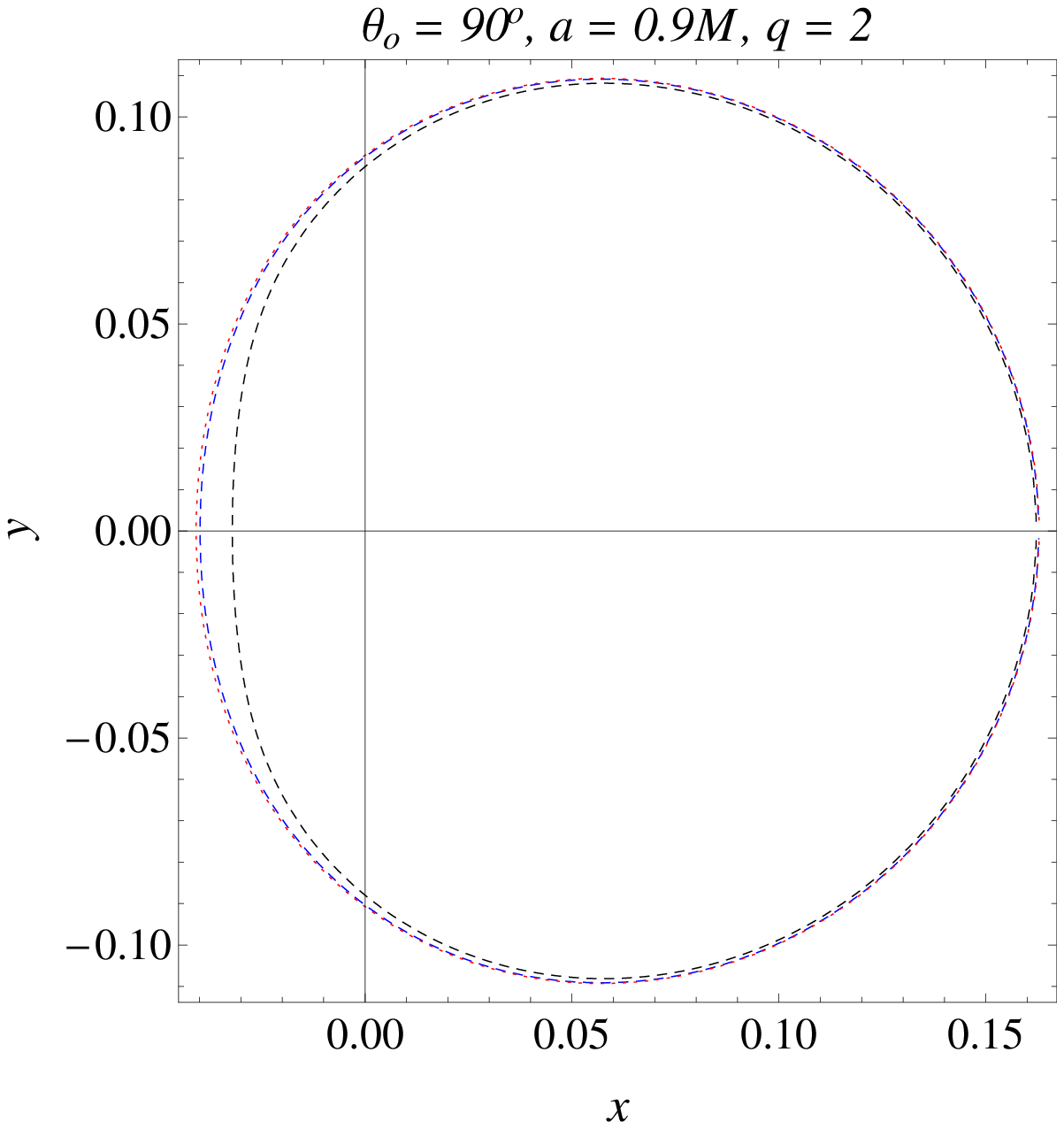}
\par\end{centering}
\caption{Shadows for the class of rotating regular black holes given by Eq. (\ref{Boyer-Lindquist}) with different values
 of $r_0$. In the second and third figures (from left to right), the influence of large values of $r_0$ decreases and
  distorts clearly the shadow. The parameters $\Lambda=10^{-3}M$, $r_{obs}=35M$, and $r_0=0.25M$, $r_0=10^{-1}M$, $r_0=10^{-2}M$ and $M=1$ are adopted in these graphics.}
\label{Shadow2}
\end{figure*}

In order to describe the shadow's silhouette it is appropriate to define the Cartesian coordinates
\begin{eqnarray}
x(r_p)&=&-2\tan\left(\frac{\alpha(r_p)}{2} \right)\sin\left(\beta(r_p) \right), 
\label{x} \\
y(r_p)&=&-2\tan\left(\frac{\alpha(r_p)}{2} \right)\cos\left(\beta(r_p) \right).
\label{y}
\end{eqnarray}
Therefore, the parametric equations (\ref{x}) and (\ref{y}) draw the shadow's silhouette. According to Fig. \ref{Shadow1}, the shadow is symmetrical to the $x$-axis. For a Kerr black hole with mass $M$,
 the silhouette depends on the rotation parameter $a$ and the observer polar angle $\theta_o$, considering an observer at
  infinity. On the other hand, for the de Sitter case, the observer is in the domain of outer communication ($r=r_{obs}$),
   but still far away from the black hole, and the silhouette depends on the cosmological constant as well. In particular,
    according to Fig. \ref{Shadow2}, the shadow will depend on the  parameters of the mass function (\ref{Mass_term})
     for regular black holes as well. As we will see, the parameter $r_0$, which generates regular metrics, can increase the distortion or the deviation from
      circularity and decrease the size of the shadow.   

As is known from the Kerr metric, the black hole shadow (its form or silhouette) is strongly sensible to the parameters $a$
 and $\theta_o$. However, for the class of regular black holes given by (\ref{Boyer-Lindquist}), we have two new
 parameters that modify the shadow: $q$ and $r_0$. Here it is shown just cases in which the rotation parameter is smaller than the black
  hole mass, i.e., $a^2<M^2$, and, in order to produce shadows compatible with M87*, it is adopted $r_0\ll M$,
   which, alongside   $a^2<M^2$, provides the spacetime structure with three horizons indicated by (\ref{Structure}).   
    As  we can see from Fig. \ref{Shadow1} and Fig. \ref{Shadow2}, for $\theta_o \neq 0$, the more rotation, the more 
     moved in the positive $x$-direction the shadow is. Such a motion points toward the rotation direction. In particular,  
      the silhouette is  deformed on the left (assuming that the rotation is from left to right) for large values of $a$, 
      $\theta_o$ and for $r_0$.  This  difference between the left  and the right sides is due to the photon orbits and the
       spacetime dragging. On the left,  photons travel in the same direction of the black hole rotation, on the right they travel
        in the opposite direction. Points on the left and on the right in the shadow's silhouette are given by $r_{p-}$ and
         $r_{p+}$, respectively. In Fig.  \ref{Shadow2}, we see the parameter $r_0$ can increase the shadow deformation for
          large values of $\theta_o$. On the other hand, small values of $\theta_o$ can produce highly symmetrical shadows
           even for large values of $a$. Assuming that the M87* is observed at $17^{\circ}$, we note that the deviation from
            circularity is small, and, for large values of $\theta_o$, interestingly large values of $r_0$ decrease the shadow
             size, according to Fig. \ref{Shadow2} (see the second and the third shadows).

\section{Oblateness and deviation from circularity}
In this section, two observables are built, oblateness and deviation from circularity. But following the Event Horizon Telescope collaboration, I will use the latter in order to compute an upper bound on the GUP parameter. 

\begin{figure}[b]
\begin{centering}
\includegraphics[scale=0.35]{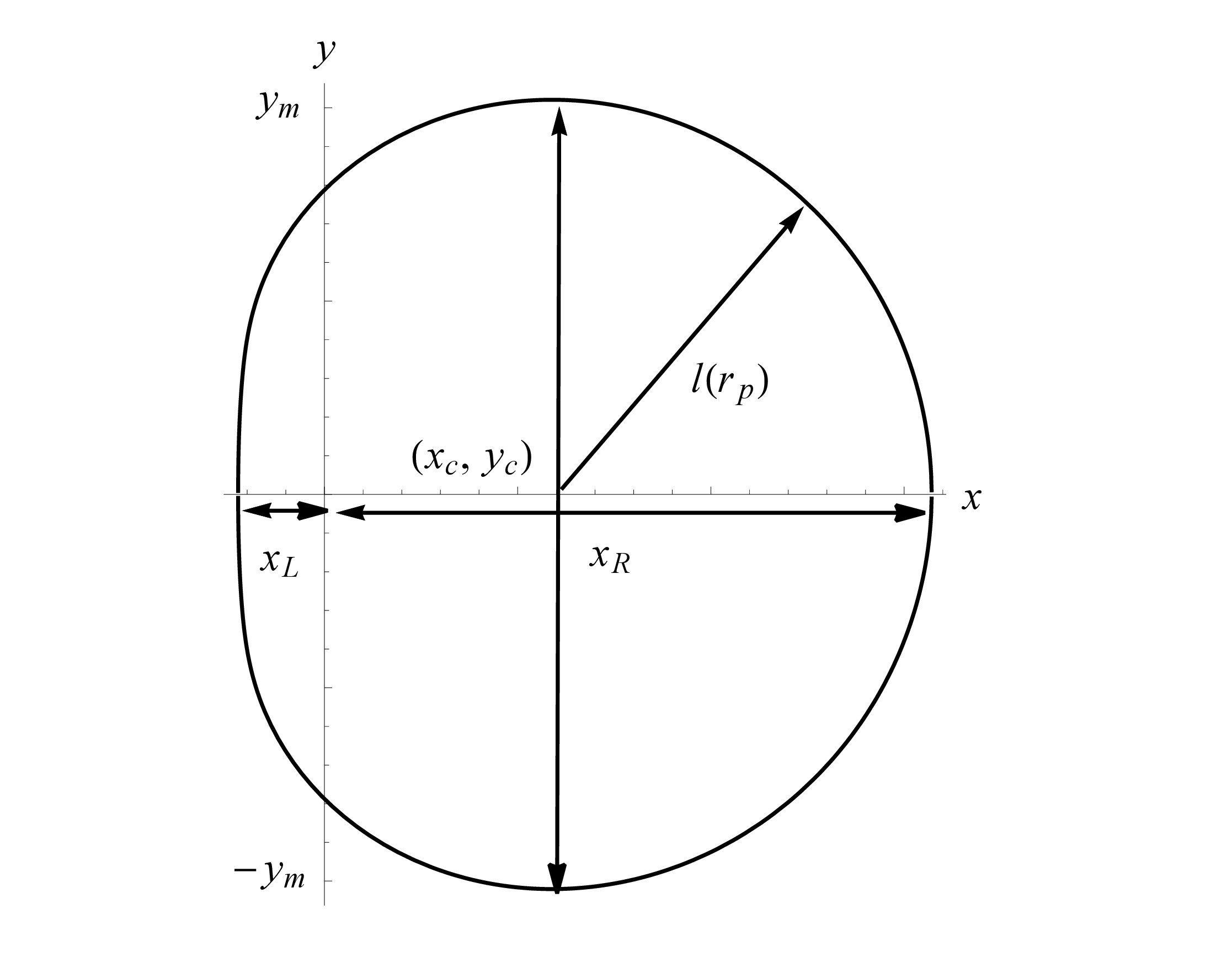}
\par\end{centering}
\caption{A schematic representation of the black hole shadow. The distances $x_L$ and $x_R$ indicate the celestial coordinate $x$ of the most negative and most positive values assumed by that coordinate. $\pm y_m$ are extremal points in the $y$-axis. The new point $(x_c,y_c)$ is constructed in order to define the shadow's radius $l(r_p)$ and calculate the deviation from circularity.}
\label{Example}
\end{figure}

\begin{figure*}
\begin{centering}
\includegraphics[scale=0.48]{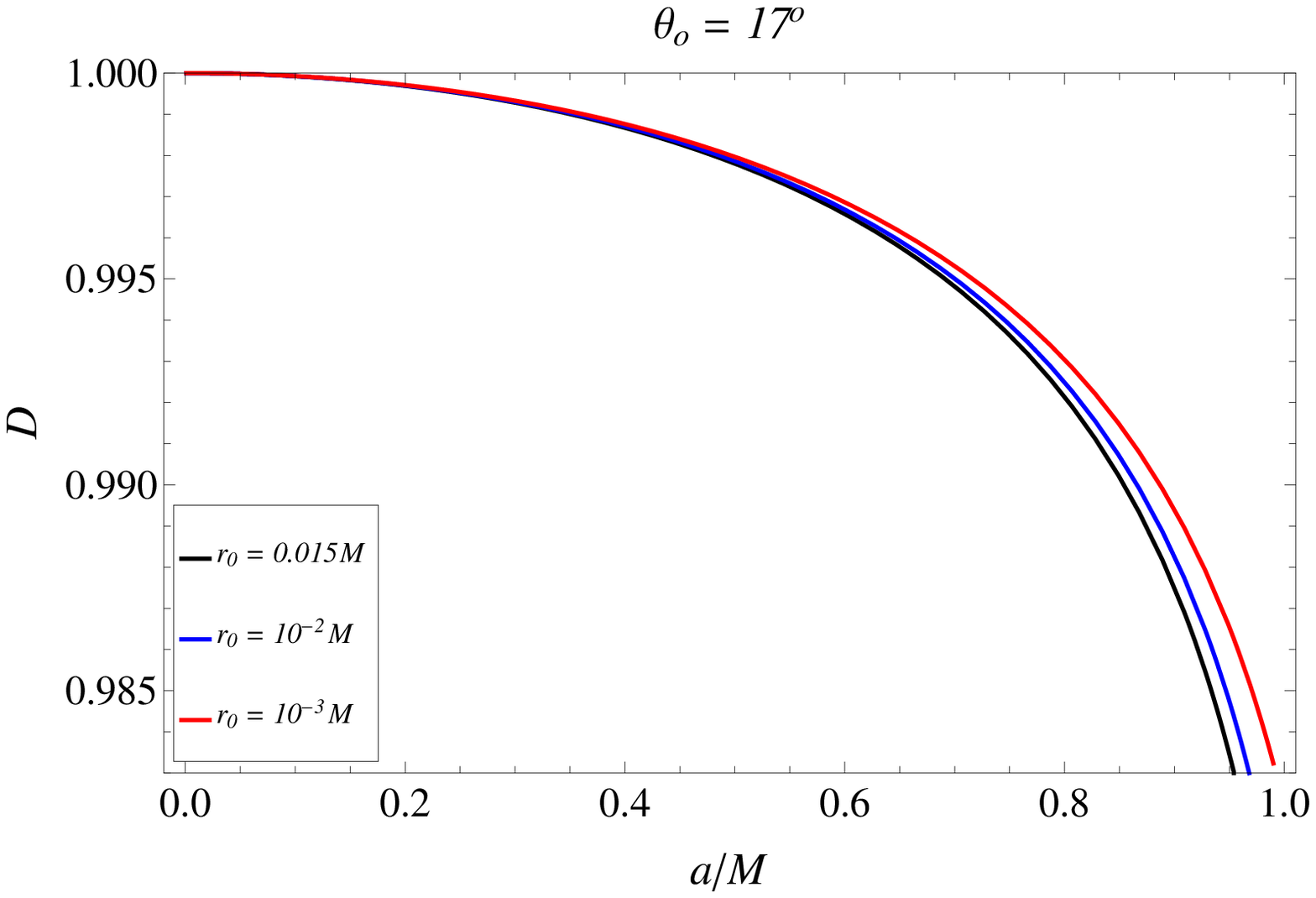}\includegraphics[scale=0.48]{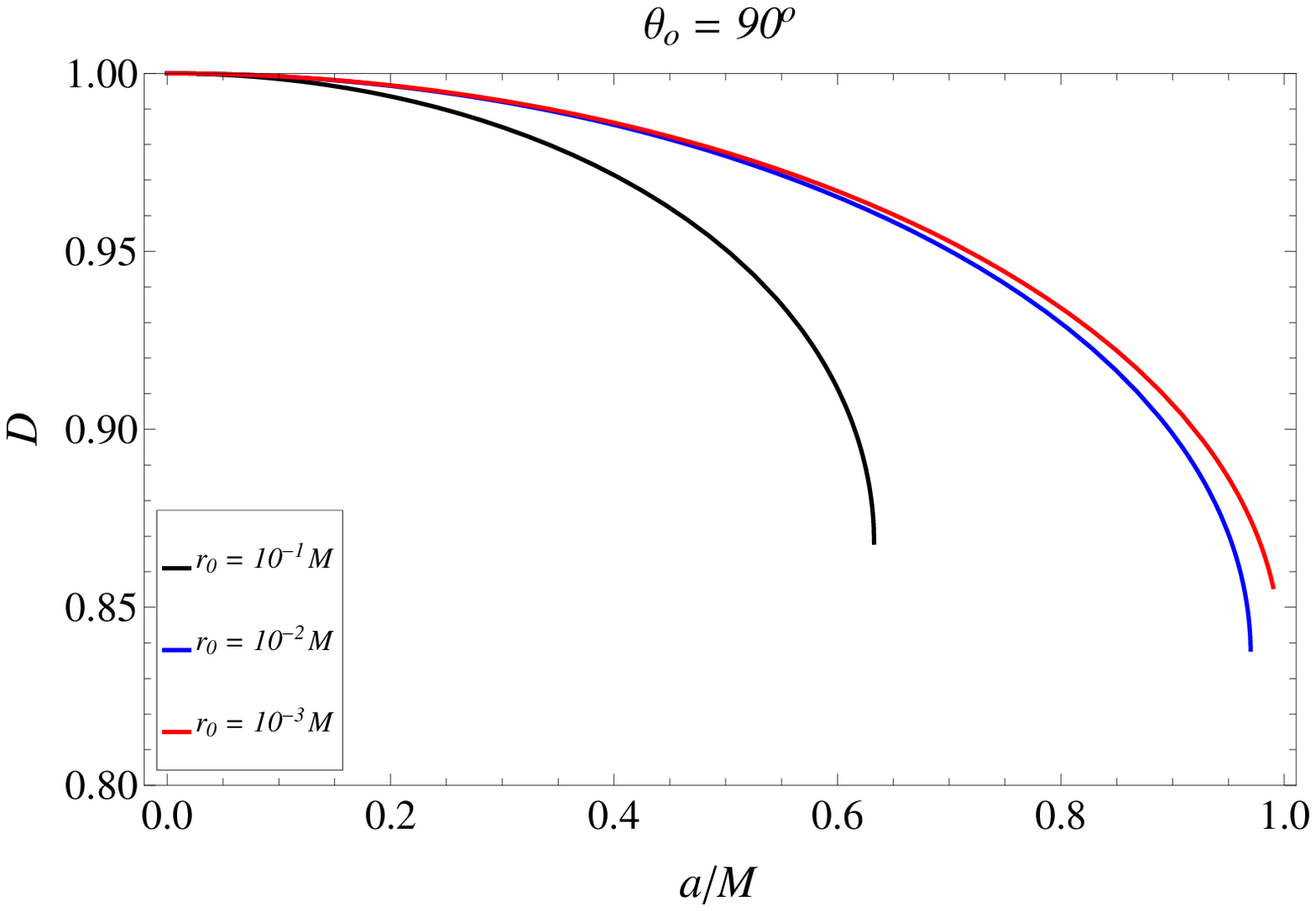}
\includegraphics[scale=0.48]{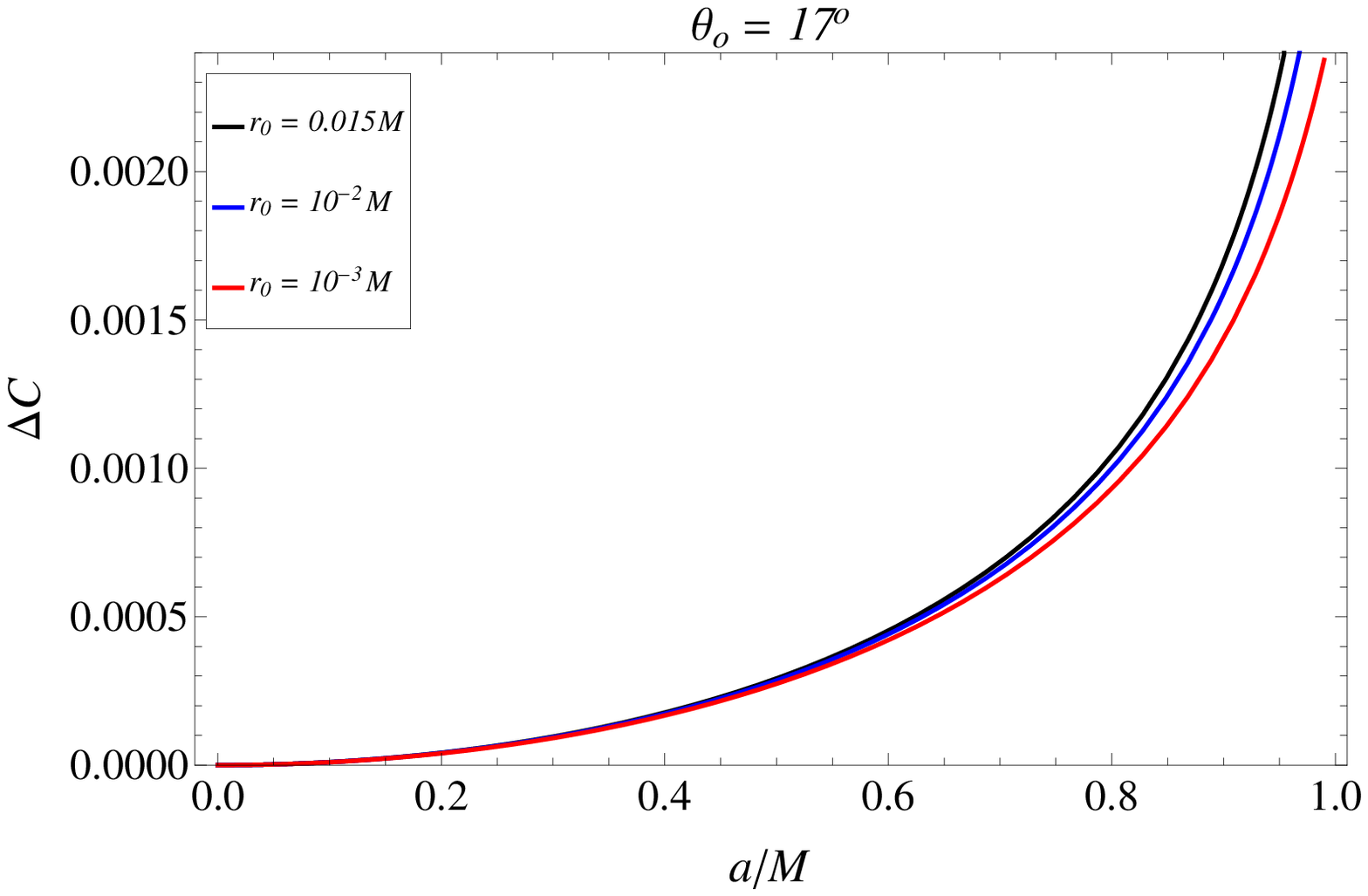}\includegraphics[scale=0.48]{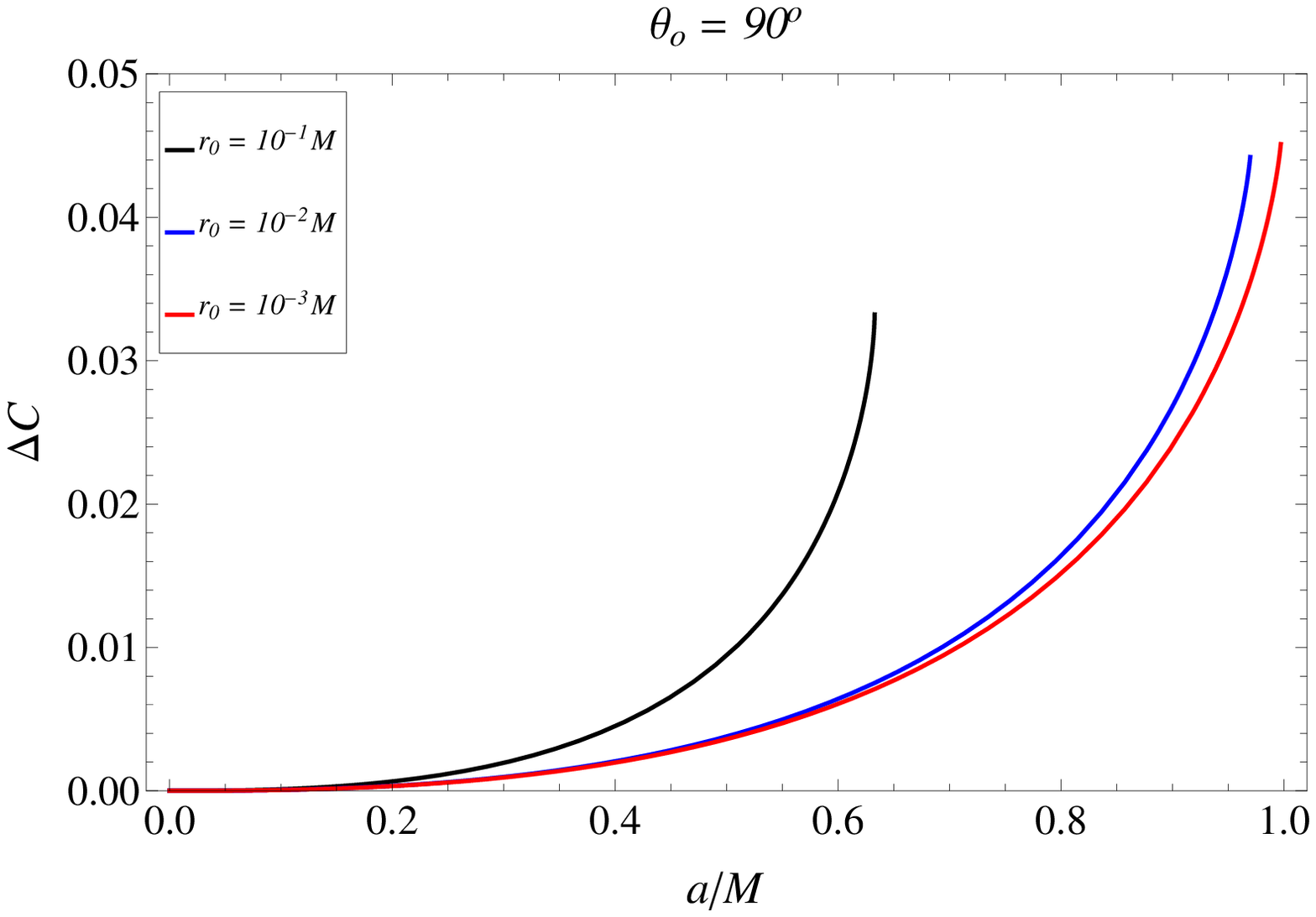}
\par\end{centering}
\caption{Oblateness $D$ and deviation from circularity $\Delta C$ for the class of rotating regular black holes given by
 Eq. (\ref{Boyer-Lindquist}). As we can see, the parameter $r_0$, that which generates regular metrics, decreases the oblateness and increases the deviation from circularity. The parameters $\Lambda=10^{-3}M$, $r_{obs}=10M$, $r_0=10^{-3}M$, and $M=1$ are adopted in these graphics.}
\label{D}
\end{figure*}

\subsection{Oblateness}
Following Refs. \cite{Bambi,Maeda,Tsupko,Kumar}, let us construct two observables for the black hole shadow. The first one indicates the difference between the $x$ and $y$ axes. Such a difference $D$ is the oblateness. The second one, which was used for the Event Horizon Telescope collaboration, provides the shadow's deviation from circularity using the root-mean-square of the shadow's radius.

In order to evaluate $D$, one adopts an approximation for the Cartesian celestial coordinates $x$ and $y$ due to the large distance of the observer, $r_{obs}=(16.8 \pm 0.8)$Mpc, which turns the value of $\alpha$ into a small quantity. Thus, Eqs. (\ref{x}) and (\ref{y}) are rewritten as     
\begin{eqnarray}
x(r_p)&\simeq&-\sin \left(\alpha(r_p)\right)\sin\left(\beta(r_p) \right),
\label{x_approx}  \\
y(r_p)&\simeq&-\sin \left(\alpha(r_p)\right)\cos\left(\beta(r_p) \right).
\label{y_approx}
\end{eqnarray}

The oblateness is simply defined as
\begin{equation}
D=\frac{\Delta x}{\Delta y}=\frac{x_R-x_L}{2y_m},
\label{Oblateness}
\end{equation}
in which
\begin{equation}
x_L=-\sin\alpha(r_{p-}) \ \ \ \ \mbox{and} \ \ \ \ x_R=\sin \alpha(r_{p+}),
\end{equation} 
where $r_{p-}$ is solution of $\sin \beta=1$, and $r_{p+}$ is given by $\sin \beta=-1$. According to Fig. (\ref{Example}), the values $x_L$ and $x_R$ (left and right) indicate maximum and minimum values assumed by the shadow silhouette on the $x$-axis. On the other hand, $y_m$ is obtained from (\ref{y_approx}) and its maximum, i.e., $\frac{dy(r_p)}{dr_{p}}=0$ leads to
\begin{equation}
\sin \alpha\frac{d}{dr_p}\cos \beta+\cos \beta \frac{d}{dr_p}\sin \alpha  = 0. 
\label{ym}
\end{equation}   
And the roots of Eq. (\ref{ym}) provide the value of $r$ at which
\begin{equation}
y_m=y(r_{p'}).
\end{equation}
The parameter $r=r_{p'}$ ($r_{p-}<r_{p'}<r_{p+}$) indicates the silhouette's maximum value on the $y$-axis. And, as we can see in all shadows shown here, $\pm y_m$ are maximum and minimum values of $y$ due to axial symmetry. Moreover, as we will see, the parameters $x_L$, $x_R$, and $y_m$ will be useful in order to construct the deviation from circularity using the root-mean-square of the shadow's radius. 

By using Eq. (\ref{Oblateness}) for members of the class of rotating regular black holes studied here, Fig. \ref{D} indicates a strong dependence between oblateness and the rotation parameter $a$. The more rotation, the more deformed the shadow is. Moreover, the parameter $r_0$, which is related to the GUP parameter, modifies $D$, decreasing the oblateness.

\subsection{Deviation from circularity}
According to the Event Horizon Telescope collaboration \cite{EHT}, a measure for the shadow deformation is provided by the deviation from circularity, which is conceived of as deviation from the root-mean-square of the radius $l(r_p)$. Following Bambi \textit{et al.} \cite{Bambi} (but here another parameterization is adopted\footnote{Bambi \textit{et al.} \cite{Bambi} use the angle defined by $l$ and the $x$-axis in order to parameterize the shadow radius.}), the shadow radius $l(r_p)$ is defined as
\begin{equation}
l(r_p)=\sqrt{(x-x_c)^2+(y-y_c)^2},
\end{equation}
with $x_c=x(r_{p'})$ and $y_c=0$ (see Fig. \ref{Example}). The average radius (root-mean-square) is given by
\begin{equation}
l_{RMS}=\sqrt{\frac{1}{(r_{p+}-r_{p-})}\int_{r_{p-}}^{r_{p+}}l(r_p)^2 dr_p}.
\end{equation}
According to the cited authors, who follow the report from the collaboration, the deviation from circularity is conceived of as the root-mean-square distance from the average radius $l_{RMS}$, that is to say, 
\begin{equation}
\Delta C = \sqrt{\frac{1}{(x_R-x_L)}\int_{r_{p-}}^{r_{p+}}\left (l(r_p) - l_{RMS} \right)^2 dr_p}.
\label{DeltaC}
\end{equation}
However, as I said, another parametrization is adopted and suggested here, demanding that $(x_R-x_L)$ is in the
 denominator of (\ref{DeltaC}). Like the oblateness, $\Delta C$ increases with the black hole rotation and the observation
  angle (see Fig. \ref{D}). The Event Horizon Telescope reported $\Delta C <0.1$ assuming a Kerr metric as the geometry
   of M87*. Here Kerr-like objects are adopted, and using the M87* parameters the deviation from circularity given by Eq.
    (\ref{DeltaC}) will provide  an upper bound on the GUP parameter.

\section{Estimating the quantum gravity parameter}
In order to constrain the GUP parameter, it is necessary to relate it to the spacetime metric. In Ref. \cite{Maluf_Neves2}, it is applied a GUP in order to calculate quantum corrections to the black hole temperature. From the GUP
\begin{equation}
\triangle x\triangle p\geq\hslash\left(1+\frac{\beta_0 l_{p}^{2}}{\hslash^{2}}\triangle p^{2}\right),
\end{equation}
where $l_{p}=\sqrt{\frac{\hslash G}{c^{3}}}\approx10^{-35}\ \text{m}$ is the Planck length, and $\beta_0$ is the so-called GUP parameter or the dimensionless quantum gravity parameter (as we can see, it is straightforward that for $\beta_0 \rightarrow 0$ we have the standard uncertainty relation), we derived the quantum-corrected temperature for the Schwarzschild black hole and, interestingly, such a temperature is the Hawking temperature for the Bardeen regular black hole (up to second order approximation in $l_p/r_+$). Thus, the Bardeen metric was interpreted as a quantum-corrected Schwarzschild black hole under the assumption that the GUP parameter could be related to a metric parameter ($r_0$, the parameter that generates regular black holes introduced in Section 2), namely,
\begin{equation}
r_0 = \frac{\beta_0^\frac{1}{2} l_p}{3}.
\label{r0}
\end{equation}
Assuming that the relation (\ref{r0}) is valid for the entire class of regular black holes, for any $q$ in Eq. (\ref{Mass_term}), whether spherical or axisymmetric black holes, then the shadow of M87* can provide an upper bound on $r_0$ and, consequently, on $\beta_0$. That is to say, the deviation from circularity (\ref{DeltaC}) depends on the spacetime metric and its parameters like $r_0$, thus, by using the relation (\ref{r0}), $\Delta C$ will depend on $\beta_0$ as well. Consequently, the upper bound on the GUP parameter can be obtained by using $\Delta C (\beta_0)$ and the constraint
provided by the Event Horizon Telescope Collaboration for the deviation from circularity.   

\begin{figure}
\begin{centering}
\includegraphics[scale=0.5]{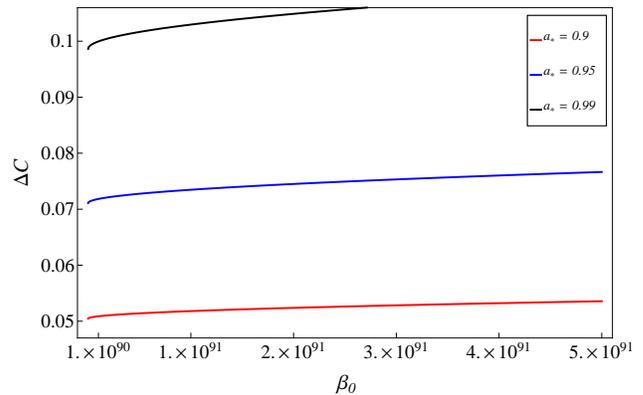}
\par\end{centering}
\caption{Deviation from circularity $\Delta C$, given by Eq. (\ref{DeltaC}), for a rotating regular black hole with M87* parameters ($M=(6.5 \pm 0.7 )\times 10^{9}M_{\odot}$, $\theta_{o}=17^{\circ}$ and $r_{obs}= (16.8 \pm 0.8)$Mpc) over the GUP parameter $\beta_0$. The reported deviation from circularity $< 10\%$, assuming the interval $a_* \lesssim 0.99$, imposes an upper bound on the GUP parameter, i.e., $\beta_0<10^{90}$. In this graphic, it is adopted $q=1$ in the mass function (\ref{Mass_term}), which renders the best upper bound on $\beta_0$, and $\Lambda\simeq 1.1\times 10^{-52}$m.}
\label{D_real}
\end{figure}

Therefore, by using $\Delta C (\beta_0)$, given by Eq. (\ref{DeltaC}), and assuming the M87* parameters, i.e., $M=(6.5 \pm 0.7 )\times 10^{9}M_{\odot}$ and $r_{obs}= (16.8 \pm 0.8)$Mpc, the deviation from circularity reported by the Collaboration, $\Delta C (\beta_0)< 0.1$, imposes
\begin{equation}
\beta_0 <10^{90},
\label{Upper bound}
\end{equation}
for $ a_* \lesssim 0.99$ (with $a_*=a/M$) and $\theta_o=17^{\circ}$, for $q=1,2$ or $q=3$ in the mass function
 (\ref{Mass_term}). As we can see in Fig. \ref{D_real}, the constraint on the deviation from circularity gives an upper
  bound on the GUP parameter, i.e., curves that satisfy  $\Delta C (\beta_0)< 0.1$ with $ a_* \lesssim 0.99$ are possible
   only for values given by Eq. (\ref{Upper bound}). On the other hand, for $a_* > 0.99$, the  deviation from circularity is
    always $\Delta C (\beta_0) >0.1$. Thus, for $\theta_o=17^{\circ}$, the parameter $a_* > 0.99$  is ruled out according
     to M87* shadow. As I pointed out in Introduction, the value $\theta_o=17^{\circ}$ is conceived of as the angle
      between the jet and the observer, assuming  that the jet direction is orthogonal to the M87* equatorial plane. The range
       for the rotation parameter presented here is in agreement with studies like \cite{Bambi}, where the Kerr metric is
        adopted in order to describe the M87* shadow. 
  
  \begin{table}
\caption{Upper bounds on the GUP parameter $\beta_0$ according to some approaches. Scanning tunneling microscope, Lamb shift and Landau Levels results are found in Ref. \cite{Das_Vagenas}. Gravitational waves result is found in Ref. \cite{Feng}, and the upper bounds from the light deflection, pulsar PRS B 1913+16 and perihelion precession were obtained in Ref. \cite{Scardigli_Casadio}. It is worth mentioning that the work on gravitational waves \cite{Feng} obtained an even better upper bound when a different GUP was applied ($< 10^{20}$, result comparable to upper bounds from quantum mechanics options).}
\label{Values of lambda}
\begin{ruledtabular}
\begin{tabular}{lcc}
     
      & $\beta_0$& \\ \hline
      \text{Scanning tunneling microscope} & $<10^{21}$ & \\ 
      \text{Lamb shift}& $<10^{36}$ &  \\ 
      \text{Landau levels}& $<10^{50}$  &     \\ 
      \text{Gravitational waves} & $<10^{60}$ & \\
      \text{Perihelion precession}&$<10^{69}$   & \\
      \text{Pulsar (PRS B 1913+16)}&$<10^{71}$   & \\
      \text{Light deflection}&$<10^{78}$   & \\
      \text{Black hole shadow}& $<10^{90}$ & \\
               
\end{tabular}
\end{ruledtabular}
\end{table}    

As we can see in Table \ref{Values of lambda}, $\beta_0<10^{90}$ is the worst value compared to other upper bounds
 provided by different approaches, like the Lamb shift, Landau levels and scanning tunneling microscope in Ref.
  \cite{Das_Vagenas}, or using light deflection, pulsar PRS B 1913+16 and perihelion precession in Ref. \cite{Scardigli_Casadio}, or from the
   gravitational waves in Ref. \cite{Feng}. According to Das and Vagenas \cite{Das_Vagenas}, the scanning tunneling
    microscope delivers the best one, $\beta_0<10^{21}$. On the other hand, like the present work, the authors of Refs.
     \cite{Scardigli_Casadio} and  \cite{Feng} present upper bounds based on gravitational phenomena, which
      rendered worse values than those provided by quantum systems. Quantum approaches have provided more stringent
       upper bounds on the GUP parameter.

\section{Final remarks}
The Event Horizon Telescope announced the first image of a black hole. According to the collaboration, the black hole
 shadow is well-described by the Kerr metric. However, it is argued that other options are still possible within the M87*
  parameters. Bambi \textit{et al.} \cite{Bambi}, for example, did not rule out a superspinar as a candidate for the
  geometry that produces the M87* shadow. In this article, a class of rotating regular black holes (thought of as a slight
   deviation from the Kerr metric) is adopted in order to describe the reported shadow.

The class of rotating regular black holes studied here presents a parameter according to which regular geometries or black
 holes without a singularity are generated. In our work \cite{Maluf_Neves2}, such a parameter from the spacetime metric
  was linked to the GUP parameter. GUPs appear in theories of quantum gravity and their parameters, which deform or
   generalize the Heisenberg principle, have been estimated. From a reported deviation from circularity of the M87*
    shadow ($<10\%$), an upper bound on the GUP parameter was computed. Alongside gravitational approaches adopted
     in order to obtain  an upper bound on the GUP parameter, the value obtained here, $\beta_0 < 10^{90}$, indicates that
      quantum options  to estimate such a parameter are better alternatives than the gravitational options like gravitational
       waves, light  deflection, pulsar PRS B 1913+16, perihelion precession, and  the black hole shadow studied here.
        In general, quantum approaches provide more stringent upper bounds on the GUP parameter to date.

\section*{Acknowledgments}
This study was financed in part by the Coordenação de Aperfeiçoamento de Pessoal de Nível Superior-Brasil (CAPES)-Finance Code 001. I thank Vilson Zanchin's research group for comments and suggestions during a seminar presentation at the UFABC. I also thank an anonymous referee for important comments and suggestions.


\begin{thebibliography}{99}

\bibitem{EHT}K. Akiyama \textit{et al.} (The Event Horizon Telescope Collaboration), Astrophys. J. Lett. \textbf{875}, L1 (2019). arXiv:\ 1906.11238

\bibitem{EHT2}K. Akiyama \textit{et al.} (The Event Horizon Telescope Collaboration), Astrophys. J. Lett. \textbf{875}, L6 (2019). arXiv:\ 1906.11243

\bibitem{Bambi}C. Bambi, K. Freese, S. Vagnozzi, and L. Visinelli, Phys. Rev. D \textbf{100},  044057 (2019). arXiv:1904.12983

\bibitem{Synge}J. L. Synge, Mon. Not. R. Astron. Soc. \textbf{131}, 463 (1966).

\bibitem{Bardeen2}J. M. Bardeen, Timelike and null geodesics in the Kerr metric, in \textit{Black Holes}, edited by C. DeWitt and B. DeWitt (Gordon and Breach, New York, 1973), p. 215.

\bibitem{Zakharov}A. F. Zakharov, F. De Paolis, G. Ingrosso, and A. A. Nucita, Astron. Astrophys. \textbf{442}, 795 (2005).

\bibitem{Vries}A. de Vries, Classical Quantum Gravity \textbf{17}, 123 (2000).

\bibitem{Perlick}V. Perlick, O. Yu. Tsupko, and G. S. Bisnovatyi-Kogan, Phys. Rev. D \textbf{97}, 104062 (2018). arXiv:1804.04898

\bibitem{Grenzebach}A. Grenzebach, V. Perlick, and C. Lämmerzahl, Phys. Rev. D \textbf{89}, 124004 (2014). arXiv:1403.5234

\bibitem{Ovgun}A. Övgün, I. Sakalli, J. Saavedra,  J. Cosmol. Astropart. Phys. \textbf{10}, 041 (2018). arXiv:1807.00388

\bibitem{Eiroa1}L. Amarilla and E. F. Eiroa, Phys. Rev. D \textbf{85}, 064019 (2012). arXiv:1112.6349

\bibitem{Eiroa2}E. F. Eiroa and C. M. Sendra, Eur. Phys. J. C  \textbf{78}, 91 (2018). arXiv:1711.08380

\bibitem{Shaikh}R. Shaikh, Phys. Rev. D \textbf{98}, 024044 (2018). arXiv:\ 1803.11422

\bibitem{Zi_Bambi}Z. Li and C. Bambi,  J. Cosmol. Astropart. Phys. \textbf{01}, 041 (2014). arXiv:1309.1606

\bibitem{Abdujabbarov}A. Abdujabbarov, M. Amir, B. Ahmedov, S.G. Ghosh, Phys. Rev. D \textbf{93}, 104004 (2016). arXiv:1604.03809

\bibitem{Amir}M. Amir and S. G. Ghosh, Phys. Rev. D \textbf{94}, 024054 (2016). arXiv:1603.06382

\bibitem{Perlick2}V. Perlick, O. Y. Tsupko, Phys. Rev. D \textbf{95}, 104003 (2017). arXiv:1702.08768

\bibitem{Cunha}P. V. P. Cunha, C. A. R. Herdeiro, Gen. Relativ. Gravit. \textbf{50}, 42 (2018). arXiv:1801.00860 

\bibitem{Roman}R. A. Konoplya, Phys. Lett. B \textbf{795}, 1 (2019). arXiv:\ 1905.00064

\bibitem{Bardeen}J. M. Bardeen, in\textit{Conference Proceedings of GR5}, Tbilisi, URSS (1968), p. 174.

\bibitem{Ansoldi}S. Ansoldi, in \textit{Conference Proceedings of BH2, Dynamics and Thermodynamics of Black Holes and Naked Singularities}, Milano, Italy (2007). arXiv:0802.0330

\bibitem{Lemos_Zanchin}J. P. S. Lemos and V. T. Zanchin, Phys. Rev. D \textbf{83}, 124005 (2011). arXiv:1104.4790

\bibitem{Dymnikova}I. G. Dymnikova, Gen. Relativ. Gravit. \textbf{24}, 235 (1992).

\bibitem{Dymnikova2} I. G. Dymnikova, Int. J. Mod. Phys. D \textbf{05}, 529 (1996).

\bibitem{Dymnikova3} I. G. Dymnikova, Int. J. Mod. Phys. D \textbf{12}, 1015 (2003). arXiv:gr-qc/0304110 

\bibitem{Bronnikov}K. A. Bronnikov, Phys. Rev. D \textbf{63}, 044005 (2001). arXiv:gr-qc/0006014

\bibitem{Hayward}S. A. Hayward, Phys. Rev. Lett. \textbf{96}, 031103 (2006). arXiv:gr-qc/0506126

\bibitem{Neves}J. C. S. Neves, Int. J. Mod. Phys. A \textbf{32}, 1750112 (2017). arXiv:1508.06701

\bibitem{Neves2}J. C. S. Neves, Phys. Rev. D \textbf{92}, \ 084015 (2015). arXiv:1508.03615

\bibitem{Various_axial}A. Smailagic, E. Spallucci, Phys. Lett. B \textbf{688}, 82 (2010). arXiv:1003.3918

\bibitem{Various_axial2}L. Modesto, P. Nicolini, Phys. Rev. D \textbf{82}, 104035 (2010).  arXiv:1005.5605

\bibitem{Various_axial3}C. Bambi, L. Modesto, Phys. Lett. B \textbf{721}, 329 (2013). arXiv:1302.6075

\bibitem{Various_axial4}B. Toshmatov, B. Ahmedov, A. Abdujabbarov, Z. Stuchlik, Phys. Rev. D \textbf{89}, 104017 (2014). arXiv:1404.6443

\bibitem{Various_axial5}M. Azreg-Ainou, Phys. Rev. D \textbf{90}, 064041 (2014). arXiv:1405.2569

\bibitem{Neves_Saa}J. C. S. Neves, A. Saa, Phys. Lett. B \textbf{734}, 44 (2014). arXiv:1402.2694

\bibitem{Wald}R. Wald, \textit{General Relativity} (University of Chicago, Chicago, 1984).

\bibitem{Beato}E. Ayon-Beato, A. Garcia,  Phys. Lett. B \textbf{493}, 149 (2000). arXiv:gr-qc/0009077

\bibitem{Maluf_Neves2}R. V. Maluf, J. C. S. Neves, Int. J. Mod. Phys. D \textbf{28}, 1950048 (2019). arXiv:1801.08872.

\bibitem{Benczik}S. Benczik, L. N. Chang, D. Minic, and T. Takeuchi, Phys. Rev. A \textbf{72}, 012104 (2005). arXiv:hep-th/0502222 

\bibitem{Das_Vagenas}S. Das, E. C. Vagenas, Phys. Rev. Lett. \textbf{101},  221301 (2008). arXiv:0810.5333

\bibitem{Das_Vagenas2}S. Das, E. C. Vagenas, Can. J. Phys. \textbf{87}, 233 (2009). arXiv:0901.1768

\bibitem{Tawfik}A. Tawfik and A. Diab, Int. J. Mod. Phys. D \textbf{23}, 1430025 (2014). arXiv:1410.0206

\bibitem{Scardigli_Casadio}F. Scardigli, R. Casadio, Eur. Phys. J. C \textbf{75}, 425 (2015). arXiv:1407.0113

\bibitem{Feng}Z. W. Feng, S. Z. Yang b, H. L. Li, X. T. Zu, Phys. Lett. B \textbf{768}, 81 (2017). arXiv:1610.08549

\bibitem{Walker}R. C. Walker, P. E. Hardee, F. B. Davies, C. Ly, and W. Junor, Astrophys. J. \textbf{855}, 128 (2018). arXiv:1802.06166.

\bibitem{Maluf_Neves1}R. V. Maluf, J. C. S. Neves, Phys. Rev. D \textbf{97}, 104015 (2018). arXiv:1801.02661

\bibitem{Neves_Saa2} J. C. S. Neves, A. Saa, \textit{Accretion of perfect fluids onto a class of regular black holes}, arXiv:1906.03718. 

\bibitem{Ceren}C. H. Bayraktar, Eur. Phys. J. Plus \textbf{133}, 377 (2018). arXiv:1806.05728

\bibitem{Planck}N. Aghanim, \textit{et al.} (Planck Collaboration), \textit{Planck 2018 results. VI. Cosmological parameters}, arXiv:1807.06209.

\bibitem{Carter}B. Carter, Phys. Rev. 174, 1559 (1968).

\bibitem{Hackmann}E.Hackmann, V.Kagramanova, J.Kunz, C.Lämmerzahl, Phys. Rev. D \textbf{81}, 044020 (2010). arXiv:1009.6117.

\bibitem{Neves_Molina}J. C. S. Neves and C. Molina, Phys. Rev. D \textbf{86}, 124047 (2012). arXiv:1211.2848 

\bibitem{Maeda}K. Hioki and K. Maeda, Phys. Rev. D \textbf{80}, 024042 (2009). arXiv:0904.3575

\bibitem{Tsupko}O. Y. Tsupko, Phys. Rev. D \textbf{95}, 104058 (2017). arXiv:\ 1702.04005.

\bibitem{Kumar}R. Kumar, S. G. Ghosh, \textit{Black hole parameters estimation from its shadow}, arXiv:1811.01260.


\end{thebibliography}
\end{document}